\documentclass[preprint,aps,draft]{revtex4}

\usepackage{graphicx}
\usepackage{dcolumn}
\usepackage{bm}
\usepackage{latexsym}

\def\E{{\cal E}}
\def\N{{\cal N}}

\begin{document}

\preprint{UTAP-454}

\title{Exactly solvable model for cosmological perturbations 
in dilatonic brane worlds}

\author{Kazuya Koyama and Keitaro Takahashi}
 
\affiliation{
Department of Physics, University of Tokyo 
7-3-1 Hongo, Bunkyo, Tokyo 113-0033, Japan
}


\begin{abstract}                
We construct a model where cosmological perturbations are 
analytically solved based on dilatonic brane worlds. 
A bulk scalar field has an exponential potential in the bulk and 
an exponential coupling to the brane tension. The bulk scalar field 
yields a power-law inflation on the brane. The exact background metric
can be found including the back-reaction of the scalar field. 
Then exact solutions for cosmological perturbations 
which properly satisfy the junction conditions on the brane are 
derived. These solutions provide us an interesting model to understand the 
connection between the behavior of cosmological perturbations on the 
brane and the geometry of the bulk. Using these solutions, the behavior of an 
anisotropic stress induced on the inflationary brane by bulk gravitational 
fields is investigated. 

\end{abstract}

\maketitle
\section{Introduction}
 The possibility that our universe is a hypersurface (brane) embedded in 
a higher-dimensional spacetime (bulk) has recently attracted  much attention
\cite{HoravaWitten}. 
Particularly, a model proposed by Randall and Sundrum has attractive features
for gravity and cosmology \cite{RSI}-\cite{review}. 
In their model, a single positive tension brane is embedded in
5-dimensional Anti-de Sitter (AdS) spacetime. Ordinary matter fields are 
assumed to be confined to the brane and only the gravity can propagate 
in the bulk. A quite interesting feature of their model is that a 4-dimensional 
gravity is recovered in the low energy limit even though the size of the bulk 
is infinite. We need no longer a compactification of the extra dimension.  
Then cosmological consequences of their model have been intensively 
investigated. In string models, the gravity enjoys the company of scalar field 
such as dilaton. Then the extension of Randall and Sundrum model to dilatonic 
brane worlds also attracts much interest
\cite{Lukas}-\cite{KobayashiKoyama}. The dilatonic brane world provides us
a new models for inflationary brane world which is often called bulk inflaton 
model \cite{KobayashiKoyamaSoda}-\cite{Soda}.
The inflation on the brane is caused by the inflaton in the bulk and the bulk 
spacetime itself is not inflating. 
It has been shown that this model indeed mimics the 4-dimensional inflation model
at low energies where the Hubble horizon on the brane $H^{-1}$ is sufficiently
longer than the curvature radius $l$ in AdS bulk . 

One of the most important quantities which should be clarified in brane world models 
is the cosmological perturbation because it provides us a possibility to 
test the brane world model observationally
\cite{Mukohyama}-\cite{KoyamaSoda2}. Unfortunately, it is quite difficult 
to find the solutions for cosmological perturbations because we should consistently 
take into account the perturbations in the bulk. Recently, Minamitsuji, Himemoto and 
Sasaki investigated a behavior of cosmological perturbations in a model with 
bulk scalar field in AdS spacetime
\cite{Minamitsuji}. They used a covariant curvature formalism and 
showed that 4-dimensional results are recovered at low energies. Thus, to predict 
signatures specific to the brane world model, we should investigate the higher 
energy effects. However, it is quite difficult to perform the calculations 
at high energies because we should treat completely a 5-dimensional problem. 
Technically, it is necessary to solve complicated coupled partial differential 
equations to find the behavior of perturbations.

Hence it is eagerly desired to construct a model where we can solve the equations 
for perturbations analytically. Recently, we have developed such a model based 
on dilatonic brane worlds \cite{KoyamaTakahashi}. 
The action for this model is given by
\begin{equation}
S = \int d^{5}x \sqrt{- g_{5}} \left( \frac{1}{2 \kappa^{2}} R 
- \frac{1}{2} \partial_{\mu} \phi \partial^{\mu} \phi - \Lambda(\phi) \right)
- \int d^{4}x \sqrt{- g_{4}} \lambda(\phi),
\end{equation}
where $\kappa^{2}$ is five-dimensional gravitational constant. 
The potential of the scalar field in the bulk and on the brane are taken to be 
exponential:
\begin{eqnarray}
\kappa^{2} \Lambda(\phi) & = & \left( \frac{\Delta}{8} + \delta \right) \lambda^{2}_{0}
e^{-2 \sqrt{2} b \kappa \phi}, \\ 
\kappa^{2} \lambda(\phi) & = & \sqrt{2} \lambda_{0} e^{- \sqrt{2} b \kappa \phi}.
\end{eqnarray}
Here $\lambda_{0}$ is the energy scale of the potential, $b$ is the dilaton coupling 
and we defined 
\begin{equation}
\Delta = 4 b^{2} - \frac{8}{3}.
\end{equation}
We assume the $Z_2$ symmetry across the brane. The bulk scalar field 
$\phi$ acts as an inflaton. For $\delta \neq 0$, the brane undergoes a power-low 
inflation. The metric for 5-dimensional spacetime is written by separable functions 
of time and extra-coordinate $y$. This is technically essential because the perturbations 
can have separable solutions which enables us to derive solutions analytically. 
It is not sufficient to derive general solutions for perturbations in the bulk. 
We should find particular solutions which satisfy the correct boundary conditions
at the brane. It is in general difficult to find such particular solutions for an 
expanding brane. But, in this model, we can successfully find such solutions 
analytically. The first aim of this 
paper is to provide exact and analytic solutions for cosmological perturbations 
in this model. 

Using the solutions for perturbations, we can address the primordial
fluctuations generated during inflation. 
In a previous paper, we calculated a spectrum of curvature perturbation 
${\cal R}_c$ by quantizing a canonical variable for the second order 
action \cite{KoyamaTakahashi}. 
We found that ,even at high energies, the effects of Kalzua-Klein (KK) modes 
are negligible at long wave-length even though the amplitude of the fluctuation 
is amplified. 
However, in the brane worlds, the curvature perturbation alone does not determine 
the cosmic microwave background (CMB) anisotropies. 
The anisotropic stress induced by bulk 
gravitational fields affects the CMB anisotropies. Indeed, it is this anisotropic 
stress which gives distinct features in CMB anisotropies in brane worlds from 
4-dimensional models \cite{Large} \cite{Koyama}. 
Thus we should specify the initial condition
for the anisotropic stress generated during inflation as well as the 
curvature perturbations. The second aim of this paper is to investigate 
the behavior of anisotropic stress in this model.  

The structure of the paper is as follows. In section II, the background spacetime 
is described. In section III, the solutions for cosmological perturbations 
are derived. Section IV is devoted to the investigation of the anisotropic
stress on the brane which is generated during inflation. 
In section V, we summarize the results. In appendix A, 5-dimensional Einstein
equations for scalar perturbations are shown. In appendix B, the procedure to derive
the solutions for scalar perturbations is presented.

\section{Background}
We first review a background solution \cite{KoyamaTakahashi}. 
For $\delta=0$, the static brane solution was found \cite{Cvetic}. 
The existence of the static brane requires tuning between bulk potential
and brane tension known as Randall-Sunrum tuning.
It has been shown that for $\Delta \leq -2$, we can avoid the presence of the naked 
singularity in the bulk and also ensure the trapping of the gravity. The reality of the 
dilaton
coupling requires $-8/3 \leq \Delta$. For $\Delta=8/3$, we recover Randall-Sundrum
solution. 
The value of $\delta$, which is not necessarily small, represents a deviation from the 
Randall-Sundrum tuning. This deviation yields an inflation on the brane.

The solution for background spacetime is found as 
\begin{eqnarray}
ds^{2} &=& e^{2 W(y)} \left(  e^{2 \sqrt{2}b \kappa \phi(t)} dy^{2} 
 - d t^{2} + e^{2 \alpha(t)} \delta_{ij} dx^{i} dx^{j} \right),\nonumber\\
\phi(t,y) &=& \phi(t) + \Xi(y).
\label{eq:background}
\end{eqnarray}
The evolution equation for background metric $\alpha$ and $\phi$ are 
given by
\begin{eqnarray}
\dot{\alpha}^2+ \sqrt{2} b \kappa \dot{\phi} \dot{\alpha} &=& 
\frac{1}{6} \kappa^2 \dot{\phi}^2 - \frac{1}{3} \lambda_0^2 
\frac{\Delta+4}{\Delta} \delta e^{-2 \sqrt{2} b \kappa \phi}, \nonumber\\
\ddot{\phi}+(3 \dot{\alpha}+ \sqrt{2} b \kappa \dot{\phi}) \dot{\phi}
&=& -4 \sqrt{2} b \kappa^{-1} \lambda_0^2 \frac{\delta}{\Delta} 
e^{-2 \sqrt{2} b \kappa \phi}. 
\label{eq:backeq}
\end{eqnarray}
The solution for $\alpha(t)$ and $\phi(t)$ are obtained as  
\begin{eqnarray}
e^{\alpha(t)} &=&
(H_0 t)^{\frac{2}{3 \Delta+8}}=
(- H \eta)^{\frac{2}{3(\Delta+2)}} ,
\label{eq:solution_alpha}
\\
e^{ \sqrt{2} b \kappa \phi(t)} &=& 
H_0 t =
\left( - H \eta \right)^{\frac{3 \Delta + 8}{3 (\Delta + 2)}}.
\label{eq:solution_phi}
\end{eqnarray}
where 
\begin{equation}
H_0 = -\frac{3 \Delta+8}{3(\Delta+2)}H, \quad 
H=- (\Delta + 2) \sqrt{- \frac{\delta}{\Delta}} \lambda_{0},
\end{equation}
and  $\eta$ is a conformal time defined by
\begin{equation}
\eta = \frac{3 \Delta+8}{3(\Delta+2)} H_0^{-\frac{2}{3 \Delta+8}} 
t^{\frac{3(\Delta+2)}{(3 \Delta+8)}}.
\end{equation}
We should notice that power-law inflation occurs on the brane for $-8/3 < \Delta < -2$.
Thus in the rest of the paper we shall assume $-8/3 < \Delta < -2$.
The solutions for $W(y)$ and $\Xi(y)$ can be written as 
\begin{equation}
e^{W(y)} = {\cal H}(y)^{\frac{2}{3(\Delta+2)}},\quad e^{\kappa \Xi(y)}=
{\cal H}(y)^{\frac{2 \sqrt{2}b}{(\Delta+2)}},
\end{equation}
where 
\begin{equation}
{\cal H}(y)= \sqrt{-1-\frac{\Delta}{8 \delta}} \sinh{H y}.
\label{eq:background_negative} 
\end{equation}
Here we assumed $\frac{\Delta}{8} + \delta < 0$.
At the location of the brane $y=y_0$
the solutions should satisfy junction conditions;
\begin{equation}
\partial_y W(y) \vert_{y=y_0}=-e^{W(y_0)-\sqrt{2}b \kappa \Xi(y_0)} 
\frac{\sqrt{2}}{6} \lambda_0,\quad 
\partial_y \Xi(y) \vert_{y=y_0}=-  e^{W(y_0)-\sqrt{2}b \kappa \Xi(y_0)}
b \kappa^{-1} \lambda_0.
\end{equation}
Then the location of the brane is determined by 
\begin{equation}
\sinh Hy_0 =\left( -1-\frac{\Delta}{8 \delta} \right)^{-1/2}.
\end{equation}

It is quite usefull to note that the above 5-dimensional solution can be obtained by 
a coordinate transformation from the metric 
\begin{equation}
ds^2 = e^{2Q(z)}(dz^2 -d \tau^2 + \delta_{i j} dx^i dx^j),\quad 
e^{\kappa \phi(z)}=e^{3 \sqrt{2} b Q(z)},
\label{eq:static}
\end{equation}
where
\begin{equation}
e^{Q(z)} = (\sinh H y_0)^{-\frac{2}{3(\Delta+2)}} (H z)^{\frac{2}{3(\Delta+2)}},
\end{equation}
by
\begin{eqnarray}
z = - \eta \sinh (H y), \nonumber\\
\tau = -\eta \cosh (H y). 
\end{eqnarray}
Because the metric Eq. (\ref{eq:static}) is simple, it is 
convenient to solve the perturbations in the bulk not directly
in Eq. (\ref{eq:background}) but in Eq. (\ref{eq:static}) for 
scalar perturbations. 

The background equations on the brane Eqs.(\ref{eq:backeq}) can be described by the 
4-dimensional Brans-Dicke theory with the action
\begin{equation}
S_{4,eff}=\frac{1}{2 \kappa_4^2} \int d^4 x \sqrt{-g_4} \left[ 
\varphi_{BD} \: ^{(4)}R-\frac{\omega_{BD}}{\varphi_{BD}}(\partial \varphi_{BD})^2 
\right] - \int d^4 x \sqrt{-g_4} V_{eff}(\varphi_{BD}),
\label{eq:BD}
\end{equation}
where 
\begin{equation}
\varphi_{BD}=e^{\sqrt{2}b \kappa \phi}, \quad 
\omega_{BD}=\frac{1}{2 b^2}, \quad 
\kappa_4^2 V_{eff}(\varphi_{BD}) =- \lambda_0^2 \delta \frac{\Delta+4}{\Delta} 
\frac{1}{\varphi_{BD}}.
\end{equation}

\section{Cosmological Perturbations}
Now let us consider cosmological perturbations in this background
spacetime. Taking appropriate gauge fixing conditions, the perturbed metric
and scalar field is give by
\begin{equation}
ds^{2} = e^{2 W(y)} \left[
e^{2 \sqrt{2} b \kappa \phi(t)}dy^2 -dt^2+e^{2 \alpha(t)}
(\delta_{ij} + h_{ij}) dx^{i} dx^{j} \right],
\end{equation}
for tensor perturbations,
\begin{equation}
ds^2=e^{2 W(y)} \left( e^{2 \sqrt{2} b \kappa \phi(t)} dy^2 
-d t^2 + e^{2 \alpha(t)} \left(2 T_i dy dx^i +
2 S_i dt dx^i+ \delta_{ij} dx^i dx^j \right) \right),
\end{equation}
for vector perturbations and 
\begin{eqnarray}
ds^2 &=& e^{2 W(y)} \left[ e^{2 \sqrt{2}b \kappa \phi(t)}(1+2 N)dy^2 
+2 A  dt dy -(1+2 \Phi )dt^2+ 
e^{2 \alpha(t)}(1-2 \Psi)\delta_{ij} dx^i dx^j) \right], \nonumber\\
\phi &=& \phi(t)+\Xi(y)+\kappa^{-1} \delta \phi,
\end{eqnarray}
for scalar perturbations, where perturbations are decomposed according to
the tensorial type of perturbations with respect to 3-space metric 
$\delta_{ij}$.
Here $S_i$ and $T_i$ are transverse vector 
($\nabla^i S_i=0$ and $\nabla^i T_i=0$) and 
$h_{ij}(y,t,x)$ is a transverse and traceless tensor($h^i_i=0,
\nabla^i h_{ij}=0$) 
where $\nabla_{i}$ is the derivative operator on 3-space metric $\delta_{ij}$.
Because each type of variables obeys an independent closed set of equations 
in the 5-dimensional Einstein equations, we derive the solutions for tensor, 
vector and scalar perturbations separately. 

\subsection{Tensor perturbations}
The evolution equation for tensor perturbations is simple.
The equation for tensor perturbation $h_{ij}(y,t,x)= h(y,t)e^{i p x}e_{ij}$ is given by
\begin{equation}
e^{2 \sqrt{2} b \kappa \phi} \left[ \ddot{h} 
+ (3 \dot{\alpha} + \sqrt{2} b \kappa \dot{\phi}) \dot{h}
+e^{-2 \alpha} p^2 h \right]
= h'' +  3 W' h',
\end{equation}
where $e_{ij}$ is a polarization tensor and dot denotes the derivative with respect 
to $t$ and prime denotes the derivative with respect to $y$. 
The junction condition for $h(y,t)$ is imposed as 
\begin{equation}
\partial_y h \vert_{y=y_0} =0.
\end{equation}
We can use the separation of variables to solve this equation. The 
solution for $h_{ij}$ is given by 
\begin{equation}
h_{ij} = \int dm d^{3}p \: h(m,p) \:\: f_{m}(y) g_{m}(\eta) e^{i p x} e_{ij},
\end{equation}
where
\begin{eqnarray}
f_0(y) &=& 1, \nonumber\\
g_0(\eta) &=& (- H \eta)^{-\frac{1}{\Delta+2}} \left( H^{(1)}_{-\frac{1}{\Delta+2}}
(-p \eta)
+ c(p,0) H^{(2)}_{-\frac{1}{\Delta+2}}(-p \eta) \right),
\end{eqnarray}
\begin{eqnarray}
f_m(y) &=& (\sinh Hy)^{\frac{\Delta}{2(\Delta+2)}} 
\left(P^{\mu}_{-\frac{1}{2}+i \nu}(\cosh Hy)- 
\frac{P^{\mu+1}_{-\frac{1}{2}+i \nu} (\cosh Hy_0)}{Q^{\mu+1}_{-\frac{1}{2}+i \nu} (\cosh 
Hy_0)}
Q^{\mu}_{-\frac{1}{2} + i \nu}(\cosh Hy) \right), \nonumber\\
g_m(\eta) &=& (- H \eta)^{-\frac{1}{\Delta+2}} \left( H^{(1)}_{i \nu}(-p \eta)
+ c(p,m) H^{(2)}_{i \nu}(-p \eta) \right),
\end{eqnarray}
where $P^{\alpha}_{\beta}$ and $Q^{\alpha}_{\beta}$ are associated Legendre functions,
$H^{(1)}_{\alpha}$ and $H^{(2)}_{\alpha}$ are Hunkel functions and 
\begin{equation}
\mu = -\frac{\Delta}{2 ( \Delta+2)}, \quad
\nu= \sqrt{\frac{m^2}{H^2}- \frac{1}{(\Delta+2)^2}}.
\end{equation}

The coefficients $h(p,m)$ and $c(p,m)$ are so far arbitrary and 
these are determined if one specifies the initial conditions 
and boundary conditions in the bulk.
The canonical variables for the second-order perturbed action for tensor perturbation is 
given by
\begin{equation}
\varphi=\frac{1}{2 \kappa} h.
\label{eq:tensor}
\end{equation}
Then the second-order perturbed five-dimensional
action for the tensor perturbation is given by
\begin{equation}
\delta S^{(T)}=\frac{1}{2}\int dy dt d^3x e^{3 W(y)} e^{ \sqrt{2} b \kappa \phi(t)}
e^{3 \alpha(t)}(e^{-2 \sqrt{2} b \kappa \phi(t)} \varphi^{'2}- \dot{\varphi}^2
-e^{-2 \alpha(t)} p^2 \varphi^2).
\end{equation}
It should be noted that the modes with $0 < m < -H/(\Delta+2)$ is not normalizable.
Thus the normalizable modes have a mass gap and the continuous massive modes start
from $m=-H/(\Delta+2)$;
\begin{equation}
m \geq - \frac{H}{\Delta+2}.
\end{equation}

\subsection{Vector perturbations}
The calculations of the vector perturbations in this model are similar
with those given by Ref.\cite{vector}.
In order to solve the Einstein equation for vector perturbations, 
it is convenient to define a variable
\begin{equation}
V_i = S_i' -\dot{T}_i =V e_i,
\end{equation}
where $e_i$ is the polarization vector. 
Then the Einstein equations of $(0,i)$ and $(i,y)$ components are given by 
\begin{eqnarray}
e^{2 \alpha -2 \sqrt{2}b \kappa \phi}(V'+3 W' V ) &=& p^2 S, \nonumber\\
e^{2 \alpha} (\dot{V}+(5 \dot{\alpha}-\sqrt{2} b \kappa \dot{\phi}) V)
&=& p^2 T,
\end{eqnarray}
where we expand the variables by $e_i e^{i p x}$. 
The $(i,i)$ component is given by
\begin{equation}
\dot{S}+(3 \dot{\alpha}+\sqrt{2} b \kappa \dot{\phi}) S
= e^{-2 \sqrt{2}b \kappa \phi}(T'+3 W' T').
\end{equation}
We can easily find a master variable from which all 
perturbations are constructed; 
\begin{eqnarray}
S &=& e^{-3W -3 \alpha - \sqrt{2}b \kappa \phi} \Omega', \nonumber\\
T &=& e^{-3W -3 \alpha + \sqrt{2}b \kappa \phi} \dot{\Omega}, \nonumber\\
V &=& e^{-3W -5 \alpha + \sqrt{2}b \kappa \phi} p^2 \Omega. 
\end{eqnarray}
If $\Omega$ satisfies the evolution equation,
\begin{equation}
e^{2 \sqrt{2}b \kappa \phi} \left[ 
\ddot{\Omega}-(3 \dot{\alpha} -\sqrt{2}b \kappa \dot{\phi}) \dot{\Omega}
+e^{-2 \alpha} p^2 \right]
= \Omega'' -3 W' \Omega',
\end{equation}
the 5-dimensional Einstein equation is automatically satisfied. 
The junction condition for perturbations are imposed as 
\begin{equation}
V \vert_{y=y_0} = T \vert_{y=y_0}=0.
\end{equation}
Thus the master variable should satisfy 
\begin{equation}
\Omega \vert_{y=y_0}=0.
\end{equation}
The solution that satisfies the junction condition is obtained as 
\begin{equation}
\Omega(t,x,y) = \int dm d^{3}p \: \Omega(m,p) \:\: u_{m}(y) v_{m}(t) e^{i p x}
\end{equation}
where
\begin{eqnarray}
u_0(y) &=&  \int^y_{y_0} e^{3 W(y')} dy' ,\nonumber\\
v_0(\eta) &=& (- H \eta)^{\frac{1}{\Delta+2}} \left( H^{(1)}_{-\frac{1}{\Delta+2}}
(-p \eta)
+ c(p,0) H^{(2)}_{-\frac{1}{\Delta+2}}(-p \eta) \right),
\end{eqnarray}
\begin{eqnarray}
u_m(y) &=& (\sinh Hy)^{\frac{\Delta+4}{2(\Delta+2)}} 
\left(P^{\mu}_{-\frac{1}{2}+i \nu}(\cosh Hy)- 
\frac{P^{\mu}_{-\frac{1}{2}+i \nu} (\cosh Hy_0)}{Q^{\mu}_{-\frac{1}{2}+i \nu} (\cosh 
Hy_0)}
Q^{\mu}_{-\frac{1}{2} + i \nu}(\cosh Hy) \right), \nonumber\\
v_m(\eta) &=& (- H \eta)^{\frac{1}{\Delta+2}} \left( H^{(1)}_{i \nu}(-p \eta)
+ c(p,m) H^{(2)}_{i \nu}(-p \eta) \right),
\end{eqnarray}
where 
\begin{equation}
\mu = -\frac{\Delta+4}{2 ( \Delta+2)}, \quad
\nu=  \sqrt{\frac{m^2}{H^2}- \frac{1}{(\Delta+2)^2}}.
\end{equation}
The second order perturbed action is also written by $\Omega$ as 
\begin{equation}
\delta S^{(V)}=\frac{1}{2} p^2 \int dy dt d^3 x e^{-3W(y)} e^{ \sqrt{2} b \kappa \phi(t)}
e^{-3 \alpha(t)}(e^{-2 \sqrt{2} b \kappa \phi(t)} \Omega^{'2}- \dot{\Omega}^2
-e^{-2 \alpha(t)} p^2 \Omega^2).
\end{equation}
From this action, we can determine the normalization of the perturbations.
It should be noted that the 0-mode solution is not normalizable for vector 
perturbations. 

\subsection{Scalar perturbation}
The scalar perturbations are more complicated than tensor and vector perturbations
due to the existence of the scalar field in the bulk. Unlike a maximally symmetric 
bulk spacetime, we cannot find a master variable for scalar perturbations and this
causes the difficulty to solve the perturbations. Fortunately, in our background
spacetime, there is a simple coordinate system in the bulk, that is Eq. (\ref{eq:static}).
Thus we can use the metric Eq. (\ref{eq:static}) to find general solutions for 
perturbations 
in the bulk. In Ref \cite{scalar}, 
it was shown that there are variables which make the equations for
$N, A, \Phi, \Psi,$ and $\delta \phi$ in the bulk to be diagonalized. Then by performing 
a coordinate transformation, it is easy to find general solutions for perturbations
in our background spacetime. However, we should proceed further to derive solutions 
for perturbations. On the brane, the perturbations should satisfy the boundary conditions.
In terms of the metric perturbations, the boundary conditions are given by
\begin{eqnarray}
\Psi' \vert_{y=y_0}&=&- W' (N -\sqrt{2} b \delta \phi)
\vert_{y=y_0}, \nonumber\\
\Phi' \vert_{y=y_0}&=& W' (N -\sqrt{2} b \delta \phi)
\vert_{y=y_0}, \nonumber\\
\delta \phi' \vert_{y=y_0} &=& 3 \sqrt{2} b 
W' (N -\sqrt{2} b \delta \phi)
\vert_{y=y_0}, \nonumber\\
A \vert_{y=y_0} &=& 0. 
\label{eq:junction-scalar}
\end{eqnarray}

A problem is that if we rewrite these conditions in terms of variables that make the 
bulk equations
to be diagonalized, the boundary conditions become complicated. Indeed, we will find 
that the boundary 
conditions are not diagonalized and also they effectively contain time derivatives of the 
variables. Thus unlike
vector and tensor perturbations, imposing the boundary conditions is not so easy. 
This is in contrast to the case for the static brane which sits at constant value of the 
$z$ in the coordinate Eq.(\ref{eq:static}). 
For the static brane, the boundary conditions for variables which make the 
equations in the bulk to be diagonalized also make the boundary conditions 
to be diagonalized and also they do not contain the time derivative 
of the variables. Thus, this complexity of the boundary conditions reflects the fact 
that our brane is moving in the coordinate Eq.(\ref{eq:static}). This movement of 
the brane causes the cosmological expansion on the brane. Hence, it is an essential 
part of the calculations  to find particular solutions which satisfy the boundary 
conditions at the brane. In fact, it has been recognized that 
this is the central part of the calculations of scalar cosmological perturbations. 
In general, the bulk perturbations are not written by separable functions with respect 
to a brane coordinate and a bulk coordinate. Thus there is almost no hope to find 
particular solutions analytically and it has prevented us from understanding the 
behavior of scalar perturbations on the 
cosmological brane. In our background spacetime, the bulk perturbations are
obtained analytically by separable functions and it enables us to derive the solutions 
which properly satisfy the boundary conditions on the brane. 

Now, we will describe our procedure to derive solutions. 
We first solve the perturbation in a static coordinate;
\begin{eqnarray}
ds^2 &=& e^{2 Q(z)}((1 +2 \Gamma)dz^2 -(1+2 \phi) d \tau^2 +2 G dz d \tau 
+ (1- 2 \psi) \delta_{ij} dx^i dx^j), \nonumber\\
\phi &=&\phi(t)+ \kappa^{-1} \delta \phi.
\end{eqnarray}
It is possible to find variables which make the equations 
diagonal \cite{scalar} (see Appendix A-2)
\begin{eqnarray}
\omega_c &=& \delta \phi + 3 \sqrt{2} b \psi, \nonumber\\
\omega_{\psi} &=& \Gamma - 2 \psi, \nonumber\\
\omega_{N} & =& \Gamma - \sqrt{2}b \delta \phi ,\nonumber\\
\omega_A &=& G.
\label{eq:omega}
\end{eqnarray}
The evolution equations in the bulk are obtained from 5-dimensional Einstein equation as 
\begin{eqnarray}
\Box_5 \omega_c &=& 0 ,\nonumber\\
\Box_5 \omega_{\psi} &=& 0 ,\nonumber\\
\Box_5 \omega_{N} &=& \frac{2 (\Delta+4)}{\Delta+2} \frac{1}
{z^2} \omega_N, \nonumber\\
\Box_5 \omega_{A} & =& \frac{2}{\Delta+2} \frac{1}
{z^2} \omega_A,
\end{eqnarray}
where 
\begin{equation}
\Box_5=
\frac{\partial^2}{\partial z^2} + \frac{2}{\Delta+2} 
\frac{1}{z} \frac{\partial}{\partial z}-
\left(\frac{\partial^2}{\partial \tau^2} + p^2 \right).
\end{equation}
By performing a coordinate transformation, the evolution
equation in our background spacetime can be derived 
\begin{eqnarray}
\Box_5 \omega_c &=& 0 ,\nonumber\\
\Box_5 \omega_{\psi} &=& 0 ,\nonumber\\
\Box_5 \omega_{N} &=& \frac{2 (\Delta+4)}{\Delta+2} \frac{H^2}
{(-H \eta)^2 \sinh^2 Hy} \omega_N ,\nonumber\\
\Box_5 \omega_{A} & =& \frac{2}{\Delta+2} \frac{H^2}
{(-H \eta)^2 \sinh^2 Hy} \omega_A,
\end{eqnarray}
where 
\begin{equation}
\Box_5= \left(\frac{1}{-H \eta}  \right)^2 \left(
\frac{\partial^2}{\partial y^2} + \frac{2}{\Delta+2} H 
\coth H y \frac{\partial}{\partial y} \right) -
\left(\frac{\partial^2}{\partial \eta^2}+ \frac{\Delta+4}{\Delta+2}
\frac{1}{\eta} \frac{\partial}{\partial \eta} + p^2 \right).
\end{equation}
The solutions for $\omega_i$ are given by
\begin{eqnarray}
\omega_c &=& \int dm_c \N_{c}(m_c,p) (\sinh Hy)^{\frac{\Delta}{2(\Delta+2)}} 
\left(P^{\mu}_{-\frac{1}{2}+i \nu(m_c)}(\cosh Hy)+ C_c Q^{\mu}_{-\frac{1}{2} + i \nu(m_c)}
(\cosh(Hy)) \right) \nonumber\\
&& \times (-H \eta)^{-\frac{1}{\Delta+2}}
H_{i \nu(m_c)}( - p \eta) ,\nonumber\\
\omega_{\psi} &=& \int dm_{\psi} \N_{\psi}(m_{\psi},p) 
(\sinh Hy)^{\frac{\Delta}{2(\Delta+2)}} 
\left(P^{\mu}_{-\frac{1}{2}+i \nu(m_{\psi})}(\cosh Hy)+ C_{\psi} Q^{\mu}_{-\frac{1}{2} + 
i \nu(m_{\psi})}(\cosh(Hy))  \right) \nonumber\\
&& \times (-H \eta)^{-\frac{1}{\Delta+2}} H_{i \nu(m_{\psi})}( - p \eta), \nonumber\\
\omega_{A} &=& \int dm_{A} \N_{A}(m_{A},p) 
(\sinh Hy)^{\frac{\Delta}{2(\Delta+2)}} 
\left(P^{\mu+1}_{-\frac{1}{2}+i \nu(m_{A})}(\cosh Hy)+ C_A Q^{\mu+1}_{-\frac{1}{2} + 
i \nu(m_{A})}(\cosh(Hy))  \right) \nonumber\\
&& \times (-H \eta)^{-\frac{1}{\Delta+2}} H_{i \nu(m_A)}( - p \eta), \nonumber\\
\omega_{N} &=& \int dm_{N} \N_{N}(m_{N},p) 
(\sinh Hy)^{\frac{\Delta}{2(\Delta+2)}} 
\left(P^{\mu+2}_{-\frac{1}{2}+i \nu(m_{N})}(\cosh Hy)+ C_N Q^{\mu+2}_{-\frac{1}{2} + 
i \nu(m_{N})}(\cosh(Hy))  \right) \nonumber\\
&& \times (-H \eta)^{-\frac{1}{\Delta+2}} H_{i \nu(m_N)}( - p \eta), \nonumber\\
\end{eqnarray}
where
\begin{equation}
\mu = -\frac{\Delta}{2 ( \Delta+2)}, \quad
\nu(m) =\sqrt{\frac{m^2}{H^2}- \frac{1}{(\Delta+2)^2}}.
\end{equation}
The perturbations $\Psi, \delta \phi,N,A$ and $\Phi$ are related to
$\psi, \delta \phi, \Gamma,G$ and $\phi$ by a coordinate transformation;
\begin{eqnarray}
\Psi &=& \psi, \quad
\delta \phi = \delta \phi,\quad 
\Phi = \Psi - N, \nonumber\\
N &=& \Gamma \cosh^2 Hy - \phi \sinh^2 Hy + G \sinh Hy \cosh Hy, \nonumber\\
A &=& -(-H \eta)^{\frac{3 \Delta+8}{3(\Delta+2)}} 
\left( 2(\Gamma-\phi) \cosh Hy \sinh Hy+G(\sinh^2 Hy+ \cosh^2 Hy) \right).
\end{eqnarray}
From Eq.(\ref{eq:omega}), $\psi, \delta \phi, \Gamma,G$ and $\phi$ are 
written in terms of $\omega_i$ as 
\begin{eqnarray}
\psi &=& - \frac{2}{3(\Delta+4)}(\omega_{\psi}-\omega_N-\sqrt{2}b \omega_c), \nonumber\\
\delta \phi &=& \frac{2 \sqrt{2}b}{\Delta+4}
\left( \omega_{\psi}-\omega_N+ \frac{\sqrt{2}}{3b} \omega_c \right) , \nonumber\\
\Gamma &=& \frac{4}{3(\Delta+4)} (\omega_N+\sqrt{2}b \omega_c+3b^2 \omega_{\psi}),
\nonumber\\
\phi &=& \psi - \Gamma.
\end{eqnarray}
Then the general solutions for the metric perturbations in the bulk are 
obtained as 
\begin{eqnarray}
\Psi &=& - \frac{2}{3(\Delta+4)}(\omega_{\psi}-\omega_N-\sqrt{2}b \omega_c),\nonumber\\
\delta \phi &=& \frac{2 \sqrt{2} b}{\Delta+4}(\omega_{\psi}-\omega_N+ 
\frac{\sqrt{2}}{3 b} \omega_c ), \nonumber\\
N &=& \frac{2}{3(\Delta+4)}(2 + 3 \sinh^2 Hy) \omega_N + \left( 
\frac{4 b^2}{\Delta +4} + \frac{2(\Delta+3)}{\Delta+4} \sinh^2 Hy 
\right) \omega_{\psi} \nonumber\\
&+& \frac{2 \sqrt{2} b}{3(\Delta+4)}(2 +3 \sinh^2 Hy) \omega_c 
+ \sinh Hy \cosh Hy \omega_A ,\nonumber\\
A &=& - (-H \eta)^{\frac{3 \Delta +8}{3(\Delta+2)}}\left[ 
2 \sinh Hy \cosh Hy \left(   \frac{2}{\Delta+4} \omega_N + 
\frac{2(\Delta+3)}{\Delta+4} \omega_{\psi}+ \frac{2 \sqrt{2} b}{\Delta+4}
\omega_c \right) \right. \nonumber\\
&+& \left.  (1+2 \sinh^2 Hy) \omega_A 
\right].
\end{eqnarray}

We should impose boundary conditions on the brane.
In terms of $\omega_i$, the junction conditions become
\begin{eqnarray}
\omega_{c}' &=& 0 ,\label{eq:jc}\\
\omega_{A} &=& - \frac{2 \cosh Hy_0 \sinh Hy_0}{1 +2 \sinh^2 Hy_0}
\left(\frac{2}{\Delta+4} \omega_{N} + \frac{2(\Delta+3)}{\Delta +4}
\omega_{\psi}+ \frac{2 \sqrt{2} b}{\Delta+4} \omega_{c}\right), \label{eq:jA} \\
\omega_{\psi}'- \omega_{N}' &=& \frac{\Delta+4}{\Delta+2}H \coth Hy_0
\left(\omega_{N}+ \frac{1}{2} \tanh Hy_0 \omega_{A} \right), \label{eq:jN} \\
\frac{2}{\Delta+4} \sinh^2 Hy_0 \omega_{N}' &+& \left( 
1+ \frac{2(\Delta+3)}{\Delta+4} \sinh^2 Hy_0 \right) \omega_{\psi}'
+\sinh H y_0 \cosh H y_0 \omega_{A}' =0, \label{eq:Jpsi}
\end{eqnarray}
where above equations should be evaluated on the brane $y=y_0$.
The variables should also satisfy the "constraint equations" which do not include
the second derivatives of the variables with respect to $t$ and $y$ 
in 5-dimensional Einstein equations, that is, 
$(t,i)$, $(y,i)$ and $(y,t)$ components of Einstein equaitons. 
Among them, the equation obtained by combining $(0,i)$ and $(y,i)$ components of the 
5-dimensional Einstein equation (see Appendix A-2 for derivation)
\begin{eqnarray}
(1 +\sinh^2 Hy) \omega_A' &+& \frac{2}{\Delta+2} H \coth Hy \omega_A 
+2 \sinh Hy \cosh Hy \omega_{\psi}' \nonumber\\
&=& -(-H \eta)\left(\cosh Hy \sinh Hy \partial_{\eta}\omega_A + 2 (1+ \sinh^2 Hy) 
\partial_{\eta}\omega_{\psi} \right),
\label{eq:constraint}
\end{eqnarray}
will be useful to find solutions. 
Because $\omega_A$ is a variable which is associated with $Z_2$ odd 
variable $A$, $\omega_{A}'$ in the junction conditions would be 
eliminated using constraint equations. 
Indeed, by projecting Eq.(\ref{eq:constraint}) on the brane, 
we find that $\omega_{A}'(y_0)$ can be rewritten in terms of $\omega_A, \omega_{\psi}'$
and $\omega_{\psi}$.  An important point is that this equation contains 
the time derivative of the variables. 
Thus effectively, the boundary conditions contain the time derivative of the 
metric perturbations. It should be noted the junction condition for $\omega_c$ 
is decoupled from other variables. This variable $\omega_c$ is the canonical variable 
for the second order action and 
it is directly related to the curvature perturbations on the brane. 
Thus we do not need to know the full solutions for perturbations in 
deriving the solutions for curvature perturbation. 
On the other hand, in order to derive the anisotropic stress
induced by bulk perturbations, we should know the solutions for all variables,
as we will observe later.  
This indicates that the anisotropic stress on the brane is complicatedly coupled 
to bulk perturbations compared with the curvature perturbation.  

We describe the procedure to derive the solutions which satisfy the 
boundary conditions and constrained equations in Appendix B. In this 
section we only show the results. 
The solutions are written as the summation of KK modes with mass $m$;
\begin{equation}
\omega_i= \int d^3 p d \tilde{m} \N(\tilde{m},p)\omega_i(\tilde{m})(y,t) e^{i p x},
\label{eq:scalarKK}
\end{equation}
where $\tilde{m}=m/H$.
For 0-mode with $m=0$, we get
\begin{eqnarray}
\omega_c(0) &=& (-H \eta)^{-\frac{1}{\Delta+2}}H_{-\frac{1}{\Delta+2}}(- p \eta), 
\nonumber\\
\omega_{\psi}(0) &=& - \frac{\sqrt{2} b}{\Delta+3} \left( 
(-H \eta)^{-\frac{1}{\Delta+2}}H_{-\frac{1}{\Delta+2}}(- p \eta) \right. \nonumber\\
&& \left. +\left(1+\frac{2(\Delta+3)}{\Delta+4}  \sinh^2 Hy \right)
(-H \eta)^{-\frac{1}{\Delta+2}}H_{- \frac{2 \Delta+5}{\Delta+2}}(- p \eta)
\right),  \nonumber\\
\omega_N(0) &=& - \frac{2 \sqrt{2}b}{\Delta+4} \sinh^2 Hy
(-H \eta)^{-\frac{1}{\Delta+2}}H_{- \frac{2 \Delta+5}{\Delta+2}}(- p \eta), \nonumber\\
\omega_A(0) &=&  \frac{4 \sqrt{2} b}{\Delta+4} \sinh Hy \cosh Hy 
(-H \eta)^{-\frac{1}{\Delta+2}}H_{- \frac{2 \Delta+5}{\Delta+2}}(- p \eta).
\end{eqnarray}
For massive modes with $m \geq -H/(\Delta+2)$
\begin{eqnarray}
\omega_c(\tilde{m}) &=& 
- \frac{\sqrt{2}(\Delta+2)}{4b}(i \nu -1)
(\sinh Hy)^{\frac{\Delta}{2(\Delta+2)}} B^{\mu}_{-\frac{1}{2}+i \nu}(\cosh Hy) 
(-H \eta)^{-\frac{1}{\Delta+2}} H_{i \nu}(- p \eta),
\nonumber\\
\omega_{\psi}(\tilde{m}) &=& 
 (\sinh Hy)^{\frac{\Delta}{2(\Delta+2)}} (-H \eta)^{-\frac{1}{\Delta+2}}
\left[   
-\frac{1}{2} B^{\mu}_{-\frac{1}{2}+ i \nu}
(\cosh Hy) H_{i \nu}(-p \eta) \right. \nonumber\\
&-& \left. \frac{1}{2} 
\left(\frac{i \nu - \frac{2 \Delta+3}{\Delta+2}}
{i \nu -\frac{1}{\Delta+2}} \right)
\left(\frac{i \nu - \frac{\Delta+1}{\Delta+2}}{i \nu - \frac{\Delta+3}{\Delta+2}}
\right)B^{\mu}_{-\frac{5}{2}+i \nu}(\cosh Hy)H_{i \nu -2}(-p \eta) 
\right],\nonumber\\
\omega_{N}(\tilde{m}) &=&  (\sinh Hy)^{\frac{\Delta}{2(\Delta+2)}} (-H 
\eta)^{-\frac{1}{\Delta+2}}
\left[   
\frac{\Delta+2}{2} 
\left( \frac{1}{i \nu -\frac{1}{\Delta+2}} \right)
B^{\mu+2}_{-\frac{1}{2}+ i \nu}
(\cosh Hy) H_{i \nu}(-p \eta)  \right. \nonumber\\
&-& \left. \frac{1}{2} 
\left( \frac{1}{i \nu -\frac{1}{\Delta+2}} \right)
\left(\frac{1}{i \nu - \frac{\Delta+3}{\Delta+2}}
\right)B^{\mu+2}_{-\frac{5}{2}+i \nu}(\cosh Hy)H_{i \nu -2}(-p \eta) 
\right],\nonumber\\
\omega_A(\tilde{m}) &=& (\sinh Hy)^{\frac{\Delta}{2(\Delta+2)}} (-H 
\eta)^{-\frac{1}{\Delta+2}}
\left[  \left( \frac{1}{i \nu -\frac{1}{\Delta+2}} \right) 
 B^{\mu+1}_{-\frac{1}{2}+ i \nu}
(\cosh Hy) H_{i \nu}(-p \eta) \right. \nonumber\\
&-& \left.  \left( \frac{1}{i \nu -\frac{1}{\Delta+2}} \right)
\left(\frac{i \nu - \frac{\Delta+1}{\Delta+2}}{i \nu - \frac{\Delta+3}{\Delta+2}}
\right)B^{\mu+1}_{-\frac{5}{2}+i \nu}(\cosh Hy)H_{i \nu -2}(-p \eta) 
\right],
\end{eqnarray}
where
\begin{equation}
B^{\alpha}_{\beta}(\cosh Hy)=P^{\alpha}_{\beta}(\cosh Hy)-
\frac{P^{\mu+1}_{-\frac{1}{2}+i \nu}(\cosh Hy_0)}{Q^{\mu+1}_{-\frac{1}{2}+ i \nu}
(\cosh Hy_0 )} Q^{\alpha}_{\beta}(\cosh Hy),
\label{eq:B}
\end{equation}
and $H_{\alpha}$ is the arbitrary combination of Hunkel functions $H^{(1)}_{\alpha}$
and $H^{(2)}_{\alpha}$.
Then the solutions for metric perturbations are derived as 
\begin{eqnarray}
\Psi(y,t,x) &=& \int d^3 p d\tilde{m} \N(\tilde{m},p) \Phi(\tilde{m})(y,t) 
e^{ipx},\nonumber\\
\delta \phi(y,t,x) &=& \int d^3 p d\tilde{m} \N(\tilde{m},p) \delta \phi(\tilde{m})(y,t) 
e^{ipx}, \nonumber\\
N(y,t,x) &=& \int d^3 p d\tilde{m} \N(\tilde{m},p) N(\tilde{m})(y,t) e^{ipx}, \nonumber\\
A(y,t,x) &=& \int d^3 p d\tilde{m} \N(\tilde{m},p) A(\tilde{m})(y,t) e^{ipx}, \nonumber\\
\Phi(y,t,x) &=& \Psi(y,t,x)-N(y,t,x).
\end{eqnarray}
where
\begin{eqnarray}
\Psi(0) &=& \frac{2 \sqrt{2} b}{3 (\Delta+4)}
\left[\frac{\Delta+4}{\Delta+3}(-H \eta)^{-\frac{1}{\Delta+2}}H_{-\frac{1}{\Delta+2}}(- p 
\eta) 
+ \frac{1}{\Delta+3} (-H \eta)^{-\frac{1}{\Delta+2}}H_{- \frac{2 \Delta+5}{\Delta+2}}(- p 
\eta)
  \right], \nonumber\\
\delta \phi(0) &=& \frac{4}{3(\Delta+4)} 
\left[\frac{\Delta+4}{4(\Delta+3)}(-H \eta)^{-\frac{1}{\Delta+2}}H_{-\frac{1}{\Delta+2}}(- 
p \eta) 
- \frac{3 b^2}{\Delta+3} (-H \eta)^{-\frac{1}{\Delta+2}}H_{- \frac{2 
\Delta+5}{\Delta+2}}(- p \eta)
  \right], \nonumber\\
N(0) &=& \sqrt{2} b \delta \phi(0), \quad A(0) = 0,
\end{eqnarray}
and
\begin{eqnarray}
\Psi(\tilde{m})(y,t) &=& -\frac{2}{3(\Delta+4)}
(-H \eta)^{-\frac{1}{\Delta+2}} (\sinh Hy)^{\frac{\Delta}{2(\Delta+2)}}
\nonumber\\
&\times& \left[ 
\frac{\Delta+2}{2} \left\{ 
-\frac{1}{i \nu -\frac{1}{\Delta+2}} B^{\mu+2}_{-\frac{1}{2}+i \nu}
(\cosh Hy) + \left(i \nu - \frac{\Delta+3}{\Delta+2} \right)
B^{\mu}_{-\frac{1}{2}+ i\nu}(\cosh Hy)
\right\} H_{i\nu}(-p \eta) \right.\nonumber\\
&+& \left. \left( 
\frac{1}{i \nu - \frac{\Delta+3}{\Delta+2}}
\right)
\left(
\frac{1}{i \nu-\frac{1}{\Delta+2}}
\right)
\left\{ 
\frac{1}{2}B^{\mu+2}_{-\frac{5}{2}+i \nu}(\cosh Hy)
-\frac{1}{2} \left(i \nu - \frac{2 \Delta+3}{\Delta+2} \right)
\left(i \nu- \frac{\Delta+1}{\Delta+2} \right) \right. \right. \nonumber\\
&\times& \left. \left.
B^{\mu}_{-\frac{5}{2}+ i \nu}(\cosh Hy)
\right\} H_{i \nu -2}(-p \eta)
\right], \nonumber\\
\delta \phi(\tilde{m})(y,t) &=& \frac{2 \sqrt{2} b}{\Delta+4}
(-H \eta)^{-\frac{1}{\Delta+2}} (\sinh Hy)^{\frac{\Delta}{2(\Delta+2)}}
\left[ 
\frac{\Delta+2}{2} \right. \nonumber\\
& \times & \left.
\left\{ 
-\frac{1}{i \nu -\frac{1}{\Delta+2}} B^{\mu+2}_{-\frac{1}{2}+i \nu}
(\cosh Hy) - \frac{1}{3b^2} \left(i \nu - \frac{\Delta}{4(\Delta+2)} \right)
B^{\mu}_{-\frac{1}{2}+ i\nu}(\cosh Hy)
\right\}  H_{i\nu}(-p \eta)  \right. \nonumber\\
&+& \left. \left( 
\frac{1}{i \nu - \frac{\Delta+3}{\Delta+2}}
\right)
\left(
\frac{1}{i \nu-\frac{1}{\Delta+2}}
\right)
\left\{ 
\frac{1}{2}B^{\mu+2}_{-\frac{5}{2}+i \nu}(\cosh Hy)
-\frac{1}{2} \left(i \nu - \frac{2 \Delta+3}{\Delta+2} \right)
\left(i \nu- \frac{\Delta+1}{\Delta+2} \right) \right. \right. \nonumber\\
&\times& \left. \left.
B^{\mu}_{-\frac{5}{2}+ i \nu}(\cosh Hy)
\right\} H_{i \nu -2}(-p \eta)
\right],\nonumber\\
N(\tilde{m})(y,t) &=& 
(-H \eta)^{-\frac{1}{\Delta+2}} (\sinh Hy)^{\frac{\Delta}{2(\Delta+2)}}
\left[ 
\frac{2(\Delta+2)}{3(\Delta+4)} \right.
\nonumber\\
&\times& 
\left\{\frac{1}{i \nu - \frac{1}{\Delta+2}}
B^{\mu+2}_{-\frac{1}{2}+i \nu}
(\cosh Hy) 
- \left(i \nu - \frac{\Delta}{4(\Delta+2)} \right)
B^{\mu}_{-\frac{1}{2}+ i\nu}
(\cosh Hy)
\right\} 
H_{i\nu}(-p \eta) \nonumber\\
&+&
\left( \frac{1}{i \nu - \frac{\Delta+3}{\Delta+2}}
\right)
\left(
\frac{1}{i \nu-\frac{1}{\Delta+2}}
\right)
\left\{ -\frac{2}{\Delta+4} 
\left( \frac{1}{3}
B^{\mu+2}_{-\frac{5}{2}+i \nu}(\cosh Hy)
\right. \right. 
\nonumber\\
&+& \left.
b^2 \left(i \nu - \frac{\Delta+1}{\Delta+2} \right)
\left(i \nu- \frac{2\Delta+3}{\Delta+2} \right) 
B^{\mu}_{-\frac{5}{2}+ i \nu}(\cosh Hy) \right) 
+
 \frac{\Delta+2}{\Delta+4}(i \nu-1)
\nonumber\\
&\times& \left.
\left(   
B^{\mu+2}_{-\frac{5}{2}+ i\nu}(\cosh Hy)- 
\left(i \nu - \frac{\Delta+1}{\Delta+2} \right)
\left(i \nu- \frac{2\Delta+3}{\Delta+2} \right)
B^{\mu}_{-\frac{5}{2}+ i\nu}(\cosh Hy) \right)
\sinh^2 Hy  
 \right\} \nonumber\\
&\times&  \left. H_{i \nu -2}(-p \eta) \right], \nonumber\\
A(\tilde{m})(y,t) &=& - (-H \eta)^{\frac{3 \Delta+8}{3(\Delta+2)}}
\left(\frac{1}{i \nu - \frac{1}{\Delta+2}} \right)
\left[  -B^{\mu+1}_{-\frac{1}{2}+ i \nu}(\cosh Hy)H_{i \nu}(-p\eta)
\right. \nonumber\\
&+& 2 \sinh Hy \cosh Hy 
\left(\frac{1}{i \nu - \frac{\Delta+3}{\Delta+2}} \right)
\left\{ 
-\frac{1}{\Delta+4}B^{\mu +2}_{-\frac{5}{2}+ i \nu}(\cosh Hy)
\right. \nonumber\\
&-& \left. \frac{\Delta+3}{\Delta+4} 
\left(i \nu - \frac{2 \Delta+3}{\Delta+2} \right)
\left(i \nu - \frac{\Delta+1}{\Delta+2} \right) 
B^{\mu}_{-\frac{5}{2}+i\nu}(\cosh Hy) 
\right \}H_{i \nu-2}(-p\eta) \nonumber\\
&+& \left.  (1-2 \cosh^2 Hy)
\left(\frac{1}{i \nu - \frac{\Delta+3}{\Delta+2}} \right)
\left(i \nu - \frac{\Delta+1}{\Delta+2}\right)
B^{\mu+1}_{-\frac{5}{2}+ i \nu}(\cosh Hy)H_{i \nu-2}(-p \eta)
 \right].
\end{eqnarray}
The solutions for massive modes  
evaluated on the brane become somewhat simple because the function
$B^{\alpha}_{\beta}$ satisfies 
\begin{equation}
B^{\mu+1}_{-\frac{1}{2}+ i \nu}(\cosh H y_0)=0.
\end{equation}
Using the equations presented in Appendix B, we get 
\begin{eqnarray}
\Psi(\tilde{m})(y_0,t) &=& \frac{1}{3}
(-H \eta)^{-\frac{1}{\Delta+2}} (\sinh Hy_0)^{\frac{\Delta}{2(\Delta+2)}}
B^{\mu}_{-\frac{1}{2}+ i \nu}(\cosh Hy_0) \nonumber\\
&& \times \left[ H_{i \nu}(-p\eta)-\frac{1}{\Delta+2} 
 \left( \frac{1}{i \nu -\frac{\Delta+3}{\Delta+2}} \right) 
H_{i\nu-2}(-p \eta) \right],\nonumber\\
\Phi(\tilde{m})(y_0,t) &=& \frac{1}{2}
(-H \eta)^{-\frac{1}{\Delta+2}}(\sinh Hy_0)^{\frac{\Delta}{2(\Delta+2)}}
B^{\mu}_{-\frac{1}{2}+ i \nu}(\cosh Hy_0) \nonumber\\
&& \times \left[ \frac{1}{3} H_{i \nu}(-p\eta)+ 
 \left( \frac{i \nu - \frac{3 \Delta+7}{3(\Delta+2)}}
{i \nu-\frac{\Delta+3}{\Delta+2}} \right) H_{i\nu-2}(-p \eta) \right], \nonumber\\
\delta \phi(\tilde{m})(y_0,t)
&=& (-H \eta)^{-\frac{1}{\Delta+2}}(\sinh Hy_0)^{\frac{\Delta}{2(\Delta+2)}}
B^{\mu}_{-\frac{1}{2}+ i \nu}(\cosh Hy_0) \nonumber\\
&& \times \left[ -\frac{\sqrt{2} (\Delta+2) }{4 b} 
\left(i \nu + \frac{2}{3(\Delta+2)} \right)
H_{i \nu}(-p\eta)+ \frac{\sqrt{2}b}{\Delta+2}
 \left( \frac{1}
{i \nu-\frac{\Delta+3}{\Delta+2}} \right) H_{i\nu-2}(-p \eta) \right].\nonumber\\
\label{massive}
\end{eqnarray}
These solutions are first main results of this paper. 
They provide us the connection between 
the behavior of the perturbations on the brane and the perturbations 
in the bulk. 

The second order action for scalar perturbations is written in terms of the 
canonical variable $\omega_c$;
\begin{equation}
\delta S^{(S)}=\frac{1}{2 \kappa^2}
\int dy dt d^3x e^{3 W(y)} e^{ \sqrt{2} b \kappa \phi(t)}
e^{3 \alpha(t)}(e^{-2 \sqrt{2} b \kappa \phi(t)} \omega_c^{'2}- \dot{\omega}_c^2
-e^{-2 \alpha(t)} p^2 \omega_c^2).
\end{equation}
This can be verified using the result for the metric Eq.(\ref{eq:static}) 
because $\omega_c$ does not change by the coordinate transformation. 
This action is the same as the second order action for tensor perturbations.
Then, the massive modes with $0<m<-H/(\Delta+2)$
are not normalizable. Thus there is also mass gap for the scalar 
perturbations. 

\section{Primordial fluctuations in the bulk inflaton model}
In the previous section, we obtained the classical solutions for 
cosmological perturbations. These perturbations properly satisfy the 
boundary conditions at the brane. However, the boundary conditions on the brane
alone do not fix the solutions completely. There remains a freedom to 
choose the "weight" $\N(\tilde{m},p)$ in the summation of KK modes
 Eq.(\ref{eq:scalarKK}). These coefficients 
are fixed once one more boundary condition in the bulk is specified.
Because the brane is inflating, it is natural to specify the boundary 
conditions for the perturbations by quantum theory in the same way as 
the usual 4-dimensional inflationary model. We have already derived the 
second order 5-dimensional action for perturbations, the quantization 
can be done within the full 5-dimensional theory. 

In the previous paper, 
we have already carried out the quantization of scalar and tensor perturbations.
It was shown that the KK modes are well suppressed at large scales even if 
the energy scale of the inflation is sufficiently higher than 
the scale of the bulk. More precisely, the bulk curvature scale and the 
Hubble constant on the brane are determined by the bulk potential and the 
deviation from the RS tuning, respectively. Thus their ratio,
\begin{equation}
r = \left| \frac{\delta}{\Delta/8 + \delta} \right|,
\label{eq:HL}
\end{equation}
characterize the behavior of perturbations. If $r$ is large, we expect 
the five dimensional nature of the perturbations become important. 
However, we showed that due to the mass gap in the KK spectrum, the 
massive KK modes are hardly excited. Thus, at large scales, the 
behavior of tensor perturbation and curvature perturbation defined by
\begin{equation}
{\cal R}_c= \frac{\dot{\alpha}}{\dot{\phi}} \omega_c,
\end{equation}
are essentially four-dimensional except for the amplitude of the 
perturbations. 

Hence, we might expect that this model cannot be distinguish from the 
usual four-dimensional inflationary model. However, in the brane world,
the curvature perturbation ${\cal R}_c$ alone does not determine
CMB anisotropies. The anisotropic stress $\delta \pi_{\E}$ 
induced by bulk gravitational fields also affects the CMB anisotropies. 
The anisotropic stress is measured by the difference between 
$\Phi$ and $\Psi$;
\begin{equation}
(\Psi-\Phi) \vert_{y=y_0} \equiv \kappa^2 e^{2 \alpha} \delta \pi_{\E}.
\label{eq:anisotropic}
\end{equation}
where $\kappa_4 =\kappa \lambda_0$. 
Then we should also determine the initial condition for $\delta \pi_{\E}$ 
during the inflation. From the five-dimensional Einstein equation
we find that $\delta \pi_E$ is related to $N$;
\begin{equation}
N \vert_{y=y_0}=\kappa^2 e^{2 \alpha} \delta \pi_{\E}.
\end{equation}
This implies that it is not sufficient to determine the behavior of the 
canonical variable $\omega_c$ but we need the solutions for all $\omega_i$.
As already observed, the boundary conditions for $\omega_i$ except for $\omega_c$
are complicated and this reflects the fact that the brane is "moving". 
Thus we expect that the anisotropic stress can have a distinguishable feature
which the curvature perturbation ${\cal R}_c$ does not possess.
Because we can derive the solutions for $N$, it is possible to
investigate the behavior of $\delta \pi_{\E}$. 

\subsection{Behavior of the canonical variable $\omega_c$}
We first review the quantization of canonical variable 
$\omega_c$. The second order action for $\omega_c$ is nothing
but the action for a 5-dimensional massless scalar field. Then the 
quantization is easily carried out. The canonical variable
$\omega_c$ can be expressed as
\begin{equation}
\kappa^{-1} 
\omega_c(t,x,z) = \int d\tilde{m} d^{3}p \left[a_{p\tilde{m}} \theta_{\tilde{m}}(y) 
\chi_{\tilde{m}}(t) e^{i p x}
+ ({\rm h.c.}) \right].
\end{equation}
Here $a_{p\tilde{m}}$ is the annihilation operator and satisfies the following
commutation relation,
\begin{equation}
\left[ a_{p\tilde{m}}, a_{p'\tilde{m}'}^{\dagger} \right] = \delta(p - p') 
\delta(\tilde{m} - \tilde{m}').
\end{equation}
The modes functions are given by
\begin{eqnarray}
\theta_{0}(y) & = & \frac{1}{\sqrt{2}} 
\left( - 1 - \frac{\Delta}{8 \delta} \right)^{- \frac{1}{2(\Delta + 2)}}
\left[ \int^{\infty}_{y_{0}} (\sinh{Hy'})^{\frac{2}{\Delta + 2}} dy' \right]^{- 
\frac{1}{2}},
\label{eq:C0} \\
\theta_{\tilde{m}}(y) & = & \sqrt{\frac{H}{2}} 
\left( - 1 - \frac{\Delta}{8 \delta} \right)^{- \frac{1}{2(\Delta + 2)}}
(|\xi|^{2} + |\zeta|^{2})^{- \frac{1}{2}} (\sinh{Hy})^{\frac{\Delta}{2(\Delta+2)}} 
B^{\mu}_{-\frac{1}{2}+i \nu}(\cosh{Hy}),
\end{eqnarray}
and 
\begin{eqnarray}
\chi_{0}(\eta) & = & \frac{\sqrt{\pi}}{2} H^{-\frac{1}{2}} (- H \eta)^{- \frac{1}{\Delta + 
2}}
H_{- \frac{1}{\Delta + 2}}^{(1)}(- p \eta), \\
\chi_{\tilde{m}}(\eta) & = & \frac{\sqrt{\pi}}{2} H^{-\frac{1}{2}} (- H \eta)^{- 
\frac{1}{\Delta + 2}}
e^{- \frac{\nu \pi}{2}} H_{i \nu}^{(1)}(- p \eta),
\end{eqnarray}
where 
\begin{eqnarray}
\mu & = & - \frac{\Delta}{2(\Delta + 2)}, \quad
\label{eq:mu}
\nu  =  \sqrt{\tilde{m} - \frac{1}{(\Delta + 2)^{2}}}, \\
\xi & = & \frac{\Gamma(i \nu)}{\Gamma(\frac{\Delta + 1}{\Delta + 2} + i \nu)}, \\
\zeta  & = & \frac{\Gamma(-i \nu)}{\Gamma(\frac{\Delta + 1}{\Delta + 2} - i \nu)}
-\frac{P^{\mu+1}_{-\frac{1}{2}+i \nu}(\cosh Hy_0)}
{Q^{\mu +1}_{-\frac{1}{2}+ i \nu}(\cosh Hy_0)} \pi e^{\mu \pi i} 
\frac{\Gamma(\frac{1}{\Delta + 2} + i \nu)}{\Gamma(1 + i \nu)}. 
\label{eq:xi}
\end{eqnarray}
Now the spectrum of the KK modes $\N(\tilde{m},p)$ is determined as 
\begin{eqnarray}
\N(0,p) &=& \kappa \frac{\sqrt{2 \pi}}{4} H^{-\frac{1}{2}}  
\left( - 1 - \frac{\Delta}{8 \delta} \right)^{- \frac{1}{2(\Delta + 2)}}
\left[ \int^{\infty}_{y_{0}} (\sinh{Hy'})^{\frac{2}{\Delta + 2}} dy' \right]^{- 
\frac{1}{2}},
\nonumber\\
\N(\tilde{m},p) &=& -\kappa \frac{\sqrt{\pi}b}{\Delta+2}\left(\frac{1}{i \nu -1} \right) 
\left( - 1 - \frac{\Delta}{8 \delta} \right)^{- \frac{1}{2(\Delta + 2)}}
(|\xi|^{2} + |\zeta|^{2})^{- \frac{1}{2}}.
\end{eqnarray}
The ratio of massive modes and massless mode increases with $r$. But
the ratio becomes constant for large $r$. In the previous paper, 
it was shown that this is caused by the existence of mass gap. And also,
the amplitude of massive modes are dumped after the horizon crossing.
Thus the spectrum of the massive modes is blue tilted, so
the contribution from massive modes becomes negligible at large scales.  
We should note that the integration over $\tilde{m}$ logarithmically
diverges. Thus we need some regularization scheme. 

\subsection{Behavior of anisotropic stress}
Now we turn to the anisotropic stress
\begin{equation}
\kappa_4^2 e^{2 \alpha} \delta \pi_{\E} =
(\Psi-\Phi) \vert_{y=y_0}.
\end{equation}
First let us consider the 0-mode. The 0-mode solution satisfies
\begin{eqnarray}
\kappa_4^2 e^{2 \alpha} \delta \pi_{\E} &=&  
N(0)=\sqrt{2}b \delta \phi(0) \nonumber\\
&=& \int d^3 p \N(0,p)\frac{4\sqrt{2} b }{3(\Delta+4)} 
\left[\frac{\Delta+4}{4(\Delta+3)}(-H \eta)^{-\frac{1}{\Delta+2}}
H_{-\frac{1}{\Delta+2}}^{(1)}(- p \eta) 
\right. \nonumber\\
&& \left. - \frac{3 b^2}{\Delta+3} (-H \eta)^{-\frac{1}{\Delta+2}}H_{- \frac{2 
\Delta+5}{\Delta+2}}^{(1)}(- p \eta)
  \right] e^{ipx}.
\end{eqnarray}
As mentioned in section II, the effective theory for background spacetime
is given by the BD theory. In the BD theory the correspondent equation is 
given by
\begin{equation}
\Psi- \Phi = \frac{\delta \varphi_{BD}}{\varphi_{BD}} = \sqrt{2} b \delta \phi.
\end{equation}
As expected, the 0-mode solution can be described by the BD theory including
anisotropic stress. At late times $-p \eta \to 0$, $N(0)$ behaves as 
$N(0) =$ const. 

The massive modes also contribute to the anisotropic stress;
\begin{eqnarray}
\kappa_4^2 e^{2 \alpha} \delta \pi_{\E}&=&  
\frac{1}{2} \int d^3 p \int_{-\frac{1}{\Delta+2}}^{\infty}
d \tilde{m} \N(\tilde{m},p) (\sinh Hy_0)^{\frac{\Delta}{2(\Delta+2)}} 
B^{\mu}_{-\frac{1}{2}+ i \nu}(\cosh Hy_0)(-H \eta)^{-\frac{1}{\Delta+2}} \nonumber\\
&& \times \left[\frac{1}{3} H_{i \nu}^{(1)}(-p\eta)
-\left( 
\frac{i \nu-\frac{3 \Delta+5}{3(\Delta+2)}}{i \nu -\frac{\Delta+3}{\Delta+2}}
\right)H_{i\nu-2}^{(1)}(-p \eta) \right] e^{i p x}. 
\end{eqnarray}
At the horizon crossing $-p \eta =1$, 
the ratio of the amplitude of massive mode to 0-mode modes has 
similar feature with the curvature perturbation. The ratio increases with
$r$, but it becomes constant for large $r$ due to the mass gap. 
Note that the term proportional to $H_{i \nu -2}^{(1)}(-p \eta)$ does not give an
additional divergence in the integration over $\tilde{m}$. 

However, the subsequent evolution of the anisotropic stress is quite 
different from the curvature perturbation. After the horizon crossing,
the 0-mode remains constant. On the other hand, the massive modes
increase as $(-p \eta)^{-(2 \Delta+5)/(\Delta+2)} \propto 
e^{-3(2 \Delta+5) \alpha / 2}$ due to the term proportional to 
$H_{i \nu-2}(-p \eta)$. Thus if $\Delta < -\frac{5}{2}$, 
the massive modes will dominate over 0-mode and it seems to leave significant 
consequences in the inflationary brane world. 
Indeed, the massive modes on the brane Eqs.(\ref{massive}) grow for 
$(-p \eta) \to 0$. Then one might worry that this indicates the 
gravitational instability of the spacetime. 
However, the physical amplitude of the anisotropic stress is 
given by
\begin{equation}
\delta \pi_{\E} \propto e^{-2 \alpha} N \vert_{y=y_0} \propto 
e^{-(6 \Delta +19) \alpha/2}, 
\end{equation}
which always decreases with time for $-2 > \Delta > -8/3$. 
The same situation occurs in the analysis of the radion 
in Randall-Sundrum de Sitter two branes. Let us consider 
two de Sitter branes in $AdS_5$ spacetime. By imposing a fine tuning 
on the tensions of two branes, the distance between two branes, the radion, 
becomes constant. However, if one considers the perturbation of the radion, the 
radion has negative mass squared. In ref \cite{radion}, the effect of 
the quantum radion was investigated. They found that the 
metric perturbations in Longitudinal gauge grow due to the 
instability of the radion. However, the physical amplitude of 
the anisotropic stress itself decays. The resolution is that 
the Longitudinal gauge is not really a good gauge. It is possible to
find a gauge where all perturbations do not grow.
Thus the growth of the 
metric perturbations does not directly imply the instability of the 
spacetime. In our case, the same arguments should be applied. 
In general, the curvature perturbation ${\cal R}_c$ is the measure
of the linear perturbation amplitude. In our background the curvature 
perturbation does not show the instability. Thus the growth 
of the metric perturbation is the artifact of the bad choice of the 
gauge. 

Let us explicitly show that we can find a suitable gauge where all
perturbations are not growing. Let us consider a 4-dimensional 
gauge transformation
\begin{equation}
\eta \to \eta -\epsilon^{\eta}, \quad x^i \to x^i - \epsilon^{, i},
\end{equation}
By choosing $\partial_{\eta} \epsilon=\epsilon^{\eta}$, the
metric and the scalar field are transformed as 
\begin{eqnarray}
ds^2 &=& e^{2 \alpha} \left(-(1+2 \tilde{\Phi})d \eta^2 + 
\left((1-2 \tilde{\Psi}) \delta_{ij}+ E_{,ij} \right) dx^i dx^j \right), \nonumber\\
\tilde{\delta \phi} &=& \delta \phi - \kappa (\partial_{\eta} \phi) \epsilon^{\eta},
\end{eqnarray}
where
\begin{eqnarray}
\tilde{\Phi} &=& \Phi - \partial_{\eta} \epsilon^{\eta}-
(\partial_{\eta} \alpha) \epsilon^{\eta}, \nonumber\\
\tilde{\Psi} &=& \Psi + (\partial_{\eta} \alpha) \epsilon^{\eta}, \nonumber\\
E  &=& - \epsilon.
\end{eqnarray}
It is a non-trivial problem to find $\epsilon^{\eta}$ that simultaneously
eliminates the growing part of the $\Phi(m)$, $\Psi(m)$ and $\delta \phi(m)$.
In this case, it is possible to find such a solution. By taking
\begin{eqnarray}
\epsilon^{\eta}(\tilde{m}) &=& -\frac{1}{2H} \left(   
\frac{1}{i \nu - \frac{\Delta+3}{\Delta+2}} \right)
(\sinh Hy_0)^{\frac{\Delta}{2(\Delta+2)}}
B^{\mu}_{-\frac{1}{2}+\i \nu}(\cosh Hy_0)
(-H \eta)^{\frac{\Delta+1}{\Delta+2}} \nonumber\\
&& \times \left(H_{i \nu-2}^{(1)}(-p \eta)+ H_{i \nu}^{(1)}(- p \eta) \right),
\end{eqnarray}
the resultant metric and the scalar field become
\begin{eqnarray}
\tilde{\Psi}(\tilde{m})&=& 
\frac{1}{3} \left( \frac{i \nu-1}{i \nu -\frac{\Delta+3}{\Delta+2}} \right)
(\sinh Hy_0)^{\frac{\Delta}{2(\Delta+2)}}
B^{\mu}_{-\frac{1}{2}+ i\nu}(\cosh Hy_0)
(-H \eta)^{-\frac{1}{\Delta+2}} H_{i \nu}^{(1)}(-p\eta), \nonumber\\
\tilde{\Phi}(\tilde{m}) &=& \frac{2}{3}
\left( \frac{i \nu-1}{i \nu -\frac{\Delta+3}{\Delta+2}} \right)
(\sinh Hy_0)^{\frac{\Delta}{2(\Delta+2)}}
B^{\mu}_{-\frac{1}{2}+ i\nu}(\cosh Hy_0)
(-H \eta)^{-\frac{1}{\Delta+2}} H_{i \nu}^{(1)}(-p\eta), \nonumber\\
\tilde{\delta \phi}(\tilde{m})&=& 
-\frac{\sqrt{2}}{4 b} (\Delta+2) \left(   
i \nu -\frac{1}{3(\Delta+2)} \right)
\left(   
\frac{i \nu-1}{i \nu- \frac{\Delta+3}{\Delta+2}}
\right) \nonumber\\
&& \times (\sinh Hy_0)^{\frac{\Delta}{2(\Delta+2)}}
B^{\mu}_{-\frac{1}{2}+ i\nu}(\cosh Hy_0)
(-H \eta)^{-\frac{1}{\Delta+2}} H_{i \nu}^{(1)}(-p\eta),\nonumber\\
\partial_{\eta}E(\tilde{m}) &=& -\epsilon^{\eta} \nonumber\\
\end{eqnarray}
At late times $-p \eta \to 0$, $E$ behaves as 
\begin{equation}
E \propto (- \eta)^{-\frac{1}{\Delta+2}} \propto e^{-3 \alpha /2} \to 0.
\end{equation}
Thus we can find a gauge where all perturbations remain
small. The anisotropic shear $E$ induced by massive modes 
decreases in an inflationary background $\Delta < -2$.
Because the amplitude of $E$ at the horizon crossing is suppressed 
due to the mass gap, the effect of the massive modes 
is always negligible. 

In summary, in the anisotropic stress, the contribution of 
the massive modes will dominate over 0-mode and this causes
the growth of metric perturbations in the Longitudinal gauge
on the brane. However, we should carefully choose the gauge
in evaluating the effect of these massive modes on 
metric perturbations. It is possible to find a "good" gauge where all 
perturbations remain small.
In this gauge, the contribution of massive modes are not 
strong enough to cause the instability on the brane and 
massive modes do not leave significant consequences.  
So we conclude that, in our bulk inflaton model, the long-wavelength 
perturbations are quite similar with the perturbations in the BD theory 
described by the 0-mode solution even for the high energy inflation.  

\section{conclusion}
 In this paper, we derived the exact and analytic solutions for 
cosmological perturbations in dilatonic brane worlds. We used 
a background spacetime where the brane undergoes a power-law 
expansion due to the bulk scalar field. The effective theory
on the brane is given by a Brans-Dicke theory. 
The interesting feature of this model is that we can derive the exact 
background solutions including the back-reaction of the bulk scalar field. 
Moreover the spacetime metric is separable with respect to the brane coordinate
and the bulk coordinate. Then it is possible to solve the cosmological
perturbations analytically. 
 
 Scalar perturbations are quite complicated because of the existence
of the bulk scalar field. We can find variables which make the 
equations in the bulk to be diagonalized. But the boundary conditions for
these variables are not diagonalized  and also they effectively
contain the time derivatives of the variables. 
The exception is the canonical variable
$\omega_c$ of the action which is related to the curvature perturbation
${\cal R}_c$ on the brane. The evolution equation for $\omega_c$ 
and the junction condition on the brane are decoupled from other
variables. 
However, if one wants to derive the solutions for all metric perturbations,
we should solve the complicated boundary conditions. Because this complexity 
is caused by the expansion of the brane, the derivation of the solutions for 
this problem is a central part of the calculations of cosmological 
perturbations in the brane world. This difficulty has prevented us from 
understanding the behavior of scalar perturbations on the brane. 
In this background, it is possible to derive the solution analytically. 
Then we found the solutions for all metric perturbations which properly 
satisfy the junction conditions on the brane.

 As an application, we have investigated the behavior of the anisotropic
stress on the brane induced by the bulk perturbations. 
We used the quantum theory for the 5-dimensional perturbations to 
determine the amplitude of the 
perturbations. As was shown in our previous paper, the massive KK modes
do not significantly contribute to the curvature perturbations even 
in the high energy inflation where the Hubble scale on the brane is
larger than the curvature scale in the bulk. In the anisotropic stress,
the contribution of the massive modes is suppressed at the horizon 
crossing. Remarkably, however, the subsequent evolution of the 
anisotropic stress is quite different from the curvature perturbations
where massive modes rapidly decays.  In the anisotropic stress, the 
contribution of massive modes seems to dominate over 0-mode after the 
horizon crossing for $\Delta < -5/2$. The difference 
comes from the junction conditions. The junction conditions effectively 
include the time derivative of the variables except for the junction 
condition for curvature perturbation. This causes the growth of the 
massive modes. However, it also causes the growth of metric perturbations 
in the Longitudinal gauge. Thus a careful choice of the gauge was required 
to discuss the effects of these massive modes. 
We found a suitable gauge where all perturbations remain small. In this gauge, 
there is an anisotropic shear which describes the contribution 
of the massive modes. It was shown that the anisotropic shear decays 
in our spacetime. Thus, the contribution of massive modes are not 
strong enough to cause the instability on the brane and 
massive modes do not leave significant consequences.  
We concluded that, in our bulk inflaton model, the perturbations are quite 
similar with the perturbations in the BD theory described by the 0-mode solution 
at large scales even for the high energy inflation.   

Our analysis indicates that the behavior of anisotropic stress 
is quite complicated compared with the curvature perturbation. 
The behavior of the curvature perturbation can be determined by 
partially solving the perturbations. But in order to determine
the anisotropic stress, we needed to derive full solutions for
perturbations. This is indeed the generic feature of the brane world
cosmological perturbations. For example, in the Randall-Sundrum model,
the behavior of the curvature perturbation is determined only by the
conservation of the energy-momentum tensor on the brane at large scales. 
But the anisotropic stress can be determined only if the gravitational
field in the bulk is completely specified. So far the analysis of this
anisotropic stress is very limited due to the difficulty of solving the 
full 5-dimensional perturbations. Our solutions would provide an interesting 
toy model for the investigation about the relation between the behavior the 
anisotropic stress on the brane and the bulk gravitational field. 
In this paper, we determine the boundary condition for perturbations 
in the bulk by quantum theory. But it is also possible to consider 
other boundary conditions. For example, it might be interesting to 
consider the boundary condition which allows the existence of dark 
radiation on the brane. Of course, in the inflationary background, 
the dark radiation does not play a role, but it is certainly important
to fully understand the relation between the geometry of the bulk and the behavior 
of anisotropic stress, because it plays an essential role in the calculation
of the CMB anisotropies in the brane worlds. For this purpose, it may be useful 
to re-derive our results using a covariant curvature formalism because the 
geometrical meanings is manifest in this formulation. We will report these 
issues in the near future \cite{future}. 

\appendix
\section{5-dimensional Einstein equation}
\subsection{5-dimensional Einstein equation}
For convenience, we present the 5-dimensional Einstein 
equation for the scalar perturbations of the metric
\begin{equation}
ds^{2} =
- e^{2 \beta(t,y)} (1 + 2 \Phi) dt^{2}
+ 2 e^{2 \beta(t,y)} A dt dy
+ e^{2 \gamma(t,y)} (1 + 2 N) dy^{2}
+ e^{2 \alpha(t,y)} (1 - 2 \Psi) dx^{i} dx_{i}.
\end{equation}

$(t,t)$ component
\begin{eqnarray}
&-& 3 e^{-2 \gamma} \Psi''
+ e^{-2 \alpha} \nabla^{2} N
- 2 e^{-2 \alpha} \nabla^{2} \Psi
\nonumber \\ 
&-& 3 e^{-2 \beta} \dot{\alpha} \dot{N}
+ 3 e^{-2 \beta} (2 \dot{\alpha} + \dot{\gamma}) \dot{\Psi}
- 3 e^{-2 \gamma} \alpha' N'
- 3 e^{-2 \gamma} (4 \alpha' - \gamma') \Psi'
+ 3 e^{-2 \gamma} \dot{\alpha} A'
\nonumber \\ 
&-& 6 e^{-2 \gamma} \left[ \alpha'' + 2 (\alpha')^{2} - \alpha' \gamma' \right] N
+ 6 e^{-2 \beta} \dot{\alpha} (\dot{\alpha} + \dot{\gamma}) \Phi
+ 3 e^{-2 \gamma} \left[ \dot{\alpha}' 
                       + \dot{\alpha} (3 \alpha' + \beta' - \gamma') \right] A
\nonumber \\ 
&=& \kappa^2 (e^{-2 \beta}(\Phi \dot{\phi^2}-\dot{\phi} \dot{\delta \phi})
-\Lambda' \delta \phi + e^{-2 \gamma}(N \phi'^2 - \phi' \delta \phi')).
\end{eqnarray}

$(t,i)$ component
\begin{eqnarray}
&-& \frac{1}{2} e^{-2 \gamma} A' +
e^{-2 \beta} \dot{N}- 2 e^{-2 \beta} \dot{\Psi}
\nonumber \\ 
&-&  e^{-2 \beta} (\dot{\alpha} - \dot{\gamma}) N
- e^{-2 \beta} (2 \dot{\alpha} + \dot{\gamma}) \Phi
- \frac{1}{2} e^{-2 \gamma} (\alpha' + 3 \beta' - \gamma') A 
\nonumber\\
&=& - \kappa^2 e^{-2 \beta}\dot{\phi} \delta \phi.
\end{eqnarray}

$(t,y)$ component
\begin{eqnarray}
&-& 3 e^{-2 \beta} \dot{\Psi}'
- 3 e^{-2 \beta} \alpha' \dot{N}
- 3 e^{-2 \beta} (\alpha' - \beta') \dot{\Psi}
- 3 e^{-2 \beta} \dot{\alpha} \Phi'
+\frac{1}{2} e^{-2 \alpha} \nabla^{2} A
\nonumber \\ 
&-& 6 e^{-2 \beta} (\dot{\alpha}' + \dot{\alpha} \alpha'
                - \dot{\alpha} \beta' - \alpha' \dot{\gamma}) \Phi
- 3 e^{-2 \gamma} \left[ \alpha'' + (\alpha')^{2} - \alpha' (\beta' + \gamma')
                  \right] A \nonumber\\
&=& \kappa^2 e^{-2 \beta} 
\left[ 
A \phi'^2 + 2 \dot{\phi} \phi' \Phi- \dot{\phi} \delta \phi' -
\phi' \delta \dot{\phi}
\right].
\end{eqnarray}

$(i,i)$ component
\begin{eqnarray}
&-& e^{-2 \beta} \ddot{N}
+ 2 e^{-2 \beta} \ddot{\Psi}
- 2 e^{-2 \gamma} \Psi''
+ e^{-2 \gamma} \Phi''
+  e^{-2 \alpha} \nabla^{2} N
- e^{-2 \alpha} \nabla^{2} \Psi
+  e^{-2 \alpha} \nabla^{2} \Phi
+ e^{-2 \gamma} \dot{A}'
\nonumber \\ 
&-& e^{-2 \beta} (2 \dot{\alpha} - \dot{\beta} + 2 \dot{\gamma}) \dot{N}
+ 2 e^{-2 \beta} (3 \dot{\alpha} - \dot{\beta} + \dot{\gamma}) \dot{\Psi}
+ e^{-2 \beta} (2 \dot{\alpha} + \dot{\gamma}) \dot{\Phi}
+ e^{-2 \gamma} (2 \alpha' + 2 \beta' - \gamma') \dot{A}
\nonumber \\ 
&-& e^{-2 \gamma} (2 \alpha' + \beta') N'
- 2 e^{-2 \gamma} (3 \alpha' + \beta' - \gamma') \Psi'
+ e^{-2 \gamma} (2 \alpha' + 2 \beta' - \gamma') \Phi'
+ e^{-2 \gamma} (2 \dot{\alpha} + \dot{\beta}) A'
\nonumber \\ 
&-& 2 e^{-2 \gamma} \left[
    2 \alpha'' + \beta'' + 3 (\alpha')^{2} + 2 \alpha' \beta'
  - 2 \alpha' \gamma' + (\beta')^{2} - \beta' \gamma' 
  \right] N
\nonumber \\ 
&+& 2 e^{-2 \beta} \left[
    2 \ddot{\alpha} + \ddot{\gamma} + 3 \dot{\alpha}^{2}
    - 2 \dot{\alpha} \dot{\beta} + 2 \dot{\alpha} \dot{\gamma}
    - \dot{\beta} \dot{\gamma} + \dot{\gamma}^{2}
    \right] \Phi
\nonumber \\ 
&+& e^{-2 \gamma} \left[
    4 \dot{\alpha}' + 2 \dot{\beta}' 
    + 2 \dot{\alpha} (3 \alpha' + \beta' + \gamma')
    + \dot{\beta} (2 \alpha' + 2 \beta' - \gamma')
    - \dot{\gamma} (2 \alpha' + \beta')
    \right] A \nonumber\\
&=& \kappa^2 \left[ -e^{-2 \beta}(\Phi \dot{\phi}^2 - \dot{\phi} 
\dot{\delta \phi}) - \Lambda' \delta \phi -e^{-2 \beta} \dot{\phi}
\phi' A + e^{-2 \gamma}(N \phi^{'2}-\phi' \delta \phi')
\right].
\end{eqnarray}

$(i \neq j)$ component
\begin{equation}
N - \Psi + \Phi=0.
\end{equation}

$(i,y)$ component
\begin{eqnarray}
&2& \Psi' - \Phi' - \frac{1}{2} \dot{A}
+ (2 \alpha' + \beta') N + (\alpha' - \beta') \Phi
- \frac{1}{2} (\dot{\alpha} + \dot{\beta} + \dot{\gamma}) A \nonumber\\
&=& \kappa^2 \phi' \delta \phi.
\end{eqnarray}

$(y,y)$ component
\begin{eqnarray}
&3& e^{-2 \beta} \ddot{\Psi}
- 2 e^{-2 \alpha} \nabla^{2} \Psi
+ e^{-2 \alpha} \nabla^{2} \Phi
+ 3 e^{-2 \beta} (4 \dot{\alpha} - \dot{\beta}) \dot{\Psi}
+ 3 e^{-2 \beta} \dot{\alpha} \dot{\Phi}
+ 3 e^{-2 \gamma} \alpha' \dot{A}
\nonumber \\ 
&-& 3 e^{-2 \gamma} (2 \alpha' + \beta') \Psi'
+ 3 e^{-2 \gamma} \alpha' \Phi'
- 6 e^{-2 \gamma} \alpha' (\alpha' + \beta') N
\nonumber \\ 
&+& 6 e^{-2 \beta} (\ddot{\alpha} + 2 \dot{\alpha}^{2} 
                - \dot{\alpha} \dot{\beta}) \Phi
+ 3 e^{-2 \gamma} \left[
    \dot{\alpha}' + \alpha' (3 \dot{\alpha} + \dot{\beta} - \dot{\gamma})
    \right] A \nonumber\\
&=& \kappa^2 \left[ 
e^{-2 \beta}(\dot{\phi} \dot{\delta \phi} - \Phi \dot{\phi}^2)-\Lambda' \delta \phi
-e^{-2 \gamma}(\phi'^2 N - \phi' \delta \phi')
\right].
\end{eqnarray}

\subsection{Equations for $\omega_i$}
In order to derive the evolution equations for $\omega_i$ 
in the bulk, it is easy to use the coordinate Eq.(\ref{eq:static}).
By combining the Einstein equations and the equation of motion 
for the scalar field, we first get the evolution
equations for metric perturbations $\psi, \phi, \Gamma, G$ and $\delta \phi$;
\begin{eqnarray}
\partial_z^2 \psi +3 (\partial_z Q) \partial_z \psi - p^2 \psi 
-\partial_{\tau}^2 \psi
&=& -2 (\partial_z^2 Q) \Gamma - 2 \kappa (\partial_z Q)(\partial_z \phi)
\delta \phi + \frac{2}{3} \kappa^{-1} \frac{d \Lambda}{d \phi} e^{2Q} 
\delta \phi, \nonumber\\
\partial_z^2 \Gamma +3 (\partial_z Q) \partial_z \Gamma - p^2 \Gamma 
-\partial_{\tau}^2 \Gamma
&=& -\left(\partial_z^2 Q+3 (\partial_z Q)^2 - \kappa^2 
(\partial_z \phi)^2 \right) \Gamma \nonumber\\
&&  -\kappa
 \left((\partial_z Q)(\partial_z \phi)
-\partial_z^2 \phi \right) \delta \phi
- \frac{1}{3} \kappa^{-1} \frac{d \Lambda}{d \phi} e^{2Q} 
\delta \phi, \nonumber\\
\partial_z^2 \delta \phi +3 (\partial_z Q) \partial_z \delta \phi - p^2 
\delta \phi 
-\partial_{\tau}^2 \delta \phi
&=& 2 \kappa (\partial_z^2 \phi) \Gamma + 2 \kappa^2 (\partial_z \phi)^2
\delta \phi + \kappa^{-2} \frac{d^2 \Lambda}{d \phi^2} e^{2Q} 
\delta \phi, \nonumber\\
\partial_z^2 G +3 (\partial_z Q) \partial_z G - p^2 G -\partial_{\tau}^2 G
&=& -3 (\partial_z^2 Q) G
\end{eqnarray}
These equations can be diagonalized using $\omega_i$ defined in 
Eqs. (\ref{eq:omega}) \cite{scalar}. Using 
\begin{equation}
\partial_z Q = \frac{2}{3 (\Delta+2)} \frac{1}{z}, \quad  
\kappa \partial_z \phi = 3 \sqrt{2}b \partial_z Q, \quad 
\frac{d \Lambda}{d \phi} e^{2Q}= -2 \sqrt{2}b \kappa
\frac{\Delta}{(\Delta+2)^2} \frac{1}{z^2},
\end{equation}
we get the evolution equations for $\omega_i$. 

The constraint equations are also easy to be derived using the metric
Eq.(\ref{eq:static}). The $(\tau,i)$ component of 
Einstein equation is given by
\begin{equation}
-\frac{1}{2}(\partial_z G + 3 (\partial_z Q)G)-2 \partial_{\tau} \psi
+\partial_{\tau} \Gamma = 0. 
\end{equation}
Rewriting these equations by $\omega_i$, we get 
\begin{equation}
-\frac{1}{2}(\partial_z \omega_A + 3 (\partial_z Q) \omega_A) 
+ \partial_{\tau} \omega_{\psi} = 0.
\end{equation}
Then performing the coordinate transformation, we get Eq.(\ref{eq:constraint}). 
The remaining two constraint equations can be derived in the same way 
or can be obtained directly in our spacetime.

\section{Derivation of solutions for scalar perturbations}

\subsection{0-mode}
In order to derive the solution the formula for the derivative of Hunkel functions 
is usefull;
\begin{eqnarray}
\frac{d}{d \eta} \left[ 
(-H \eta)^{-\frac{1}{\Delta+2}} H_{-\frac{1}{\Delta+2}}(-p \eta) \right] 
 &=& -p (-H \eta)^{-\frac{1}{\Delta+2}} H_{-\frac{\Delta+3}{\Delta+2}}(-p \eta)
 \nonumber\\
\frac{d}{d \eta} \left[ 
(-H \eta)^{-\frac{1}{\Delta+2}} H_{-\frac{2 \Delta+5}{\Delta+2}}(-p \eta) \right] 
 &=& p (-H \eta)^{-\frac{1}{\Delta+2}} H_{-\frac{\Delta+3}{\Delta+2}}(-p \eta) \nonumber\\
 &+& \frac{2(\Delta+3)}{\Delta+2} H (-H \eta)^{-\frac{\Delta+3}{\Delta+2}}
 H_{-\frac{2 \Delta+5}{\Delta+2}}(-p \eta). \nonumber\\
\end{eqnarray}
Let us find the solution for $\omega_i$ with 
\begin{equation}
\omega_c = (-H \eta)^{-\frac{1}{\Delta+2}} H_{-\frac{1}{\Delta+2}}(-p \eta).
\end{equation}
First we use the constraint equation Eq.(\ref{eq:constraint}). 
Because this equation contains the first derivative with respect 
to time, the solution for $\omega_i$ necessarily includes 
$H_{-\frac{2 \Delta+5}{\Delta+2}}(-p \eta)$ because it is only 
the choice that can eliminate $H_{-\frac{\Delta+3}{\Delta+2}}(-p \eta)$
which arises when we take the time derivative of 
$H_{-\frac{1}{\Delta+2}}(-p \eta)$.
Then we assume that $\omega_i$ contains $H_{-\frac{1}{\Delta+2}}(-p \eta)$
and $H_{-\frac{2 \Delta+5}{\Delta+2}}(-p \eta)$. Using the fact 
that $\omega_i$ should satisfy the evolution equation in the bulk,
the $y$-dependence of the variables is automatically determined. 
Then substituting the ansatz for the variables into the 
junction conditions, we can determine all coefficients except 
for an over-all normalization. It is a non-trivial check that these
solutions indeed satisfy three constraint equations. It is 
easy to verify that these solutions indeed satisfy the 
constraint equations. 

\subsection{Massive modes}
For massive modes, the following formula is useful;
\begin{eqnarray}
\frac{d}{d \eta} \left[ 
(-H \eta)^{-\frac{1}{\Delta+2}} H_{i \nu}(-p \eta) \right] 
 &=& -p (-H \eta)^{-\frac{1}{\Delta+2}} H_{i \nu -1}(-p \eta) \nonumber\\
 &+& H \left(\frac{1}{\Delta+2}+ i \nu \right)(-H \eta)^{-\frac{\Delta+3}{\Delta+2}}
 H_{i \nu}(- p \eta).
 \nonumber\\
\frac{d}{d \eta} \left[ 
(-H \eta)^{-\frac{1}{\Delta+2}} H_{-2+i \nu}(-p \eta) \right] 
 &=& p (-H \eta)^{-\frac{1}{\Delta+2}} H_{i \nu - 1}(-p \eta) \nonumber\\
 &+&  H \left(\frac{2 \Delta+5}{\Delta+2}- i \nu \right)
 (-H \eta)^{-\frac{\Delta+3}{\Delta+2}} H_{-2+i \nu}(- p \eta). \nonumber\\
\end{eqnarray}
and 
\begin{eqnarray}  
&&\frac{d}{dy} \left[ 
(\sinh Hy)^{\frac{\Delta}{2(\Delta+2)}} B^{\mu+2}_{\beta}(\cosh Hy) \right]
= H(\sinh Hy)^{\frac{\Delta}{2(\Delta+2)}} \nonumber\\
&& \times \left[-\coth Hy \frac{\Delta +4}{\Delta+2} B^{\mu+2}_{\beta} (\cosh Hy)
+\left(\beta-\frac{\Delta+4}{2(\Delta+2)} \right)
\left(\beta+\frac{3 \Delta+8}{2(\Delta+2)} \right)B^{\mu+1}_{\beta}(\cosh Hy) \right] ,
\nonumber\\
&& \frac{d}{dy} \left[ 
(\sinh Hy)^{\frac{\Delta}{2(\Delta+2)}} B^{\mu}_{\beta}(\cosh Hy) \right]
= H(\sinh Hy)^{\frac{\Delta}{2(\Delta+2)}}B^{\mu+1}_{\beta} (\cosh Hy), \nonumber\\
&& \frac{d}{dy} \left[ 
(\sinh Hy)^{\frac{\Delta}{2(\Delta+2)}} B^{\mu+1}_{\beta}(\cosh Hy) \right]
= H(\sinh Hy)^{\frac{\Delta}{2(\Delta+2)}} 
\left[
\coth Hy B^{\mu+1}_{\beta}(\cosh Hy)+B^{\mu+2}_{\beta}(\cosh Hy) \right], \nonumber\\
&& \quad =  H(\sinh Hy)^{\frac{\Delta}{2(\Delta+2)}} \nonumber\\
&& \quad \times \left[ 
\left(-\frac{2}{\Delta+2} \right)\coth Hy B^{\mu+1}_{\beta}(\cosh Hy)
+ \left(\beta+\frac{\Delta}{2(\Delta+2)} \right)
\left(\beta+ \frac{\Delta+4}{2(\Delta+2)} \right) B^{\mu}_{\beta}(\cosh Hy)
\right]. \nonumber\\
\end{eqnarray}
As for the 0-mode, the ansatz for the solutions are determined so that 
they can satisfy the constraint equation Eq.(\ref{eq:constraint}).
We asssume
\begin{eqnarray}
\omega_c &=& (-H \eta)^{-\frac{1}{\Delta+2}} (\sinh Hy)^{\frac{\Delta}{2(\Delta+2)}} 
C_c B^{\mu}_{-\frac{1}{2}+i \nu} H_{i \nu}(-p \eta) \nonumber\\
\omega_{\psi} &=& (-H \eta)^{-\frac{1}{\Delta+2}} (\sinh Hy)^{\frac{\Delta}{2(\Delta+2)}} 
 \nonumber\\
&& \times \left[ C_{\psi} B^{\mu}_{-\frac{1}{2}+ i \nu}(\cosh Hy)H_{i \nu}(-p \eta) 
+D_{\psi} B^{\mu}_{-\frac{5}{2}+i \nu}(\cosh Hy) H_{i \nu-2}
\right] \nonumber\\
\omega_{A} &=& (-H \eta)^{-\frac{1}{\Delta+2}} (\sinh Hy)^{\frac{\Delta}{2(\Delta+2)}} 
 \nonumber\\
&& \times \left[ C_{A} B^{\mu+1}_{-\frac{1}{2}+ i \nu}(\cosh Hy)H_{i \nu}(-p \eta) 
+D_{A} B^{\mu+1}_{-\frac{5}{2}+i \nu}(\cosh Hy) H_{i \nu-2}
\right] \nonumber\\
\omega_{N} &=& (-H \eta)^{-\frac{1}{\Delta+2}} (\sinh Hy)^{\frac{\Delta}{2(\Delta+2)}} 
 \nonumber\\
&& \times \left[ C_{N} B^{\mu+2}_{-\frac{1}{2}+ i \nu}(\cosh Hy)H_{i \nu}(-p \eta) 
+D_{N} B^{\mu+2}_{-\frac{5}{2}+i \nu}(\cosh Hy) H_{i \nu-2} \right]
\end{eqnarray}
where $B^{\alpha}_{\beta}$ is given by Eq.(\ref{eq:B}).
First let us consider the constraint equation Eq.(\ref{eq:constraint}).
Because this equation only contains $\omega_{\psi}$ and $\omega_A$, 
it is easy to determine the coefficients using the formula for derivatives.
We get
\begin{equation}
C_{\psi} = -\frac{1}{2} \left(i \nu - \frac{1}{\Delta+2} \right) C_A,
\quad 
D_{\psi} = \frac{1}{2} \left(i \nu - \frac{2 \Delta+3}{\Delta+2} \right) D_A,
\quad 
D_A = - \left( \frac{i \nu - \frac{\Delta+1}{\Delta+2}}
{i \nu - \frac{\Delta+3}{\Delta+2}} \right) C_A.
\end{equation}
Next we use the junction conditions. From Eq.(\ref{eq:jN}), $D_N$ is 
determined as 
\begin{equation}
D_N = -\frac{1}{2} \left( 
\frac{1}{i \nu - \frac{\Delta+3}{\Delta+2}} \right)C_A.
\end{equation}
From Eq.(\ref{eq:jA}), $C_N$ is given by
\begin{equation} 
 C_N=\frac{\Delta+2}{2} C_A.
\end{equation}
We should note that at this time, the problem becomes non-trivial because
the equation should be satisfied by the coefficients which have already determined.
The point is that, on the brane, $B^{\mu+1}_{-\frac{1}{2}+i\nu}(\cosh Hy_0)=0$. 
Using this fact, we can show that 
\begin{eqnarray}
 \sinh Hy_0  \cosh Hy_0 B^{\mu+2}_{-\frac{5}{2}+i \nu}(\cosh Hy_0)
 &=&\frac{1}{2 (i \nu-1)} \left( 
 i \nu -\frac{\Delta+1}{\Delta+2} \right) 
 \left( 
i \nu - \frac{2 \Delta+5}{\Delta+2} 
 \right) B^{\mu+1}_{-\frac{5}{2} + i \nu}(\cosh Hy_0) \nonumber\\
&& - \left(i \nu - \frac{\Delta+1}{\Delta+2}\right) \cosh^2 Hy_0 
 B^{\mu+1}_{-\frac{5}{2}+ i \nu}(\cosh Hy_0) .
\label{eq:relation}
\end{eqnarray}
Using this relation, it is shown that the junction condition Eq.({\ref{eq:jN}}) 
can be satisfied. Finally, we use Eq.({\ref{eq:jA}}). We get  
\begin{equation}
C_c=-\frac{\sqrt{2}}{4 b} (\Delta+2) \left(i \nu - \frac{1}{\Delta+2} \right) 
(i \nu -1).
\end{equation}
Here we again used Eq.(\ref{eq:relation}) in order to show that the junction 
condition is satisfied. Now all coefficients are determined except for 
an over-all normalization. It remains a task to verify that 
these solutions satisfy the 
remaining two constraint equations. In order to show that, we use the formula 
for associate Legendre function and Hankel function
\begin{eqnarray}
B^{\mu+2}_{\beta}(z)+2(\mu+1)z(z^2-1)^{-\frac{1}{2}}B^{\mu+1}_{\beta}(z)
&=&(\beta+\mu)(\beta+\mu+1) B^{\mu}_{\beta}(z), 
\label{formula} \nonumber\\
H_{\beta-1}(z)+H_{\beta+1}(z) &=& 2 \beta z^{-1}H_{\beta}(z).
\end{eqnarray}
Then after long calculations, it is possible to show that the above 
solution indeed satisfy the constraint equations. 

In order to derive the solutions for metric perturbations on the brane,
we used Eq. ({\ref{formula}}) and the following equations;
\begin{eqnarray}
&& \left(i \nu - \frac{2 \Delta+3}{\Delta+2} \right)
\left(i \nu - \frac{\Delta+1}{\Delta+2} \right)
B^{\mu}_{-\frac{5}{2}+ i\nu}(\cosh Hy_0) \nonumber\\
&=&\left(i \nu - \frac{1}{\Delta+2} \right)
\left\{ 
2(i \nu-1)\sinh^2 Hy_0 + \left(i \nu - \frac{\Delta+3}{\Delta+2} \right)
\right\}
B^{\mu}_{-\frac{1}{2}+i\nu}(\cosh Hy_0),
\end{eqnarray}
\begin{eqnarray}
B^{\mu+2}_{-\frac{5}{2}+i\nu}(\cosh Hy_0)
= \left(i \nu - \frac{1}{\Delta+2} \right)
\left\{ 
\left(i \nu + \frac{1}{\Delta+2} \right)
+2 (i \nu-1) \sinh^2 Hy_0
\right\} B^{\mu}_{-\frac{1}{2}+ i\nu}(\cosh Hy_0), \nonumber\\
\end{eqnarray}
which can be derived using $B^{\mu+1}_{-\frac{1}{2}+i \nu}
(\cosh Hy_0)=0$.

\end{document}